\pdfoutput=1

\documentclass[11pt,twoside,a4paper,cmspaper,final,collab]{cms-tdr}
\def\svnVersion{50552}\def\cmsCernNo{2011-033}\def\cmsCernDate{\today}\def\cmsMessage{Submitted to the Journal of High Energy Physics}

\begin{document}\cmsNoteHeader{SUS-10-004}

\hyphenation{had-ron-i-za-tion}
\hyphenation{cal-or-i-me-ter}
\hyphenation{de-vices}
\RCS$Revision: 50551 $
\RCS$HeadURL: svn+ssh://alverson@svn.cern.ch/reps/tdr2/papers/SUS-10-004/trunk/SUS-10-004.tex $
\RCS$Id: SUS-10-004.tex 50551 2011-04-15 21:41:26Z alverson $
\cmsNoteHeader{SUS-10-004} 
\title{Search for new physics with same-sign isolated dilepton events with jets and missing transverse energy at the LHC}

\address[neu]{Northeastern University}
\address[fnal]{Fermilab}
\address[cern]{CERN}
\author[cern]{The CMS Collaboration}

\date{\today}

\newcommand{\met}{$E_{T}^{\rm miss}$\hspace*{.10ex}}
\newcommand{\RelIso}{$RelIso$\hspace*{1.0ex}}

\abstract{
The results of searches for new physics in events with two same-sign
isolated leptons, hadronic jets, and missing transverse energy in the final state
are presented.
The searches use an integrated luminosity of
35~pb$^{-1}$
of $\Pp\Pp$ collision data
at a centre-of-mass energy of 7 TeV
collected by the CMS experiment at the LHC.
The observed numbers of events agree with the standard model
predictions, and no evidence for new physics is found.
To facilitate the interpretation of our data in a broader range of new physics scenarios,
information on our event selection, detector response, and efficiencies is provided.
}

\hypersetup{%
pdfauthor={CMS Collaboration},%
pdftitle={Search for new physics with same-sign isolated dilepton events with jets and missing transverse energy at the LHC},%
pdfsubject={CMS},%
pdfkeywords={CMS, physics, software, computing}}

\maketitle 

\section{Introduction}
\label{sec:intro}

Events with same-sign isolated lepton pairs from hadron collisions are very rare in the standard model (SM) but
appear very naturally in many new physics
scenarios.
In particular, they have been proposed as signatures of
supersymmetry (SUSY)~\cite{Barnett:1993ea,Guchait:1994zk,Baer:1995va},
universal extra dimensions~\cite{Cheng:2002ab},
pair production of $T_{5/3}$ (a fermionic partner of the top quark)~\cite{Contino:2008hi},
heavy Majorana neutrinos~\cite{Almeida:1997em},
and
same-sign top-pair
resonances as predicted in theories with warped extra dimensions~\cite{Han:2009}.
In this paper we describe searches for new physics
with same-sign isolated dileptons ($\Pe\Pe$, $\Pe\mu$, $\mu\mu$,
$\Pe\tau$, $\mu\tau$, and $\tau\tau$), missing transverse
energy (\met), and hadronic jets.
Our choice of signal regions is driven by two simple observations.
First, astrophysical evidence for dark matter~\cite{dm1} suggests that we concentrate on final states with \met .
Second, observable new physics signals with large cross sections are likely to be produced
by strong interactions,
and we thus expect
significant hadronic activity in conjunction with the two same-sign leptons.
Beyond these simple guiding principles, our searches are as independent of detailed features of
new physics models as possible.
The results are based on
a data sample corresponding to an integrated luminosity
of 35 pb$^{-1}$ collected in $\Pp\Pp$ collisions
at a centre-of-mass energy of 7 TeV
by the Compact Muon Solenoid (CMS) experiment at the Large
Hadron Collider (LHC) in 2010.

This paper is organized as follows.  The CMS detector is briefly
described
in Section~\ref{sec:detector}.  The reconstruction of leptons,
\met, and jets at CMS is summarized in Section~\ref{sec:objects}.
Section~\ref{sec:sr} describes our search regions.
We perform separate searches based on
leptonic and hadronic triggers in order to cover a wider region in the parameter space of new physics.
Electron and muon triggers allow for searches that require less hadronic energy in the event, while
hadronic triggers allow inclusion of lower transverse momentum ($p_T$) electrons and muons, as well as hadronic $\tau$
decays in the final state.
The dominant backgrounds for all three searches are estimated from data,
as discussed in Section~\ref{sec:bkg}.
Systematic uncertainties on the predicted number of signal events
and results
of these searches are discussed in
Sections~\ref{sec:systematics} and~\ref{sec:results}.
We conclude with a discussion on how to use our results to constrain a wide variety of
new physics models in Section~\ref{sec:discussion}.

\section{The CMS Detector}
\label{sec:detector}
A right-handed coordinate system is employed by the CMS experiment, with the origin at the nominal interaction point,
 the $x$-axis pointing to the centre of the LHC, and the $y$-axis pointing up (perpendicular to the LHC plane).
The polar angle $\theta$ is measured from the positive
 $z$-axis and the azimuthal angle $\phi$ is measured in the $xy$ plane.
The pseudorapidity is defined as $\eta=-\ln{[\tan{(\frac{\theta}{2})}]}$.

The central feature of the CMS apparatus is a superconducting solenoid,
of 6~m internal diameter, 13~m in length, providing an axial field of
3.8~T. Within the field volume are several particle detection systems
which each feature a cylindrical geometry, covering the full azimuthal range
from $0 \le \phi \le 2\pi$.  Silicon pixel and strip tracking detectors
provide measurements of charged particle trajectories and extend to
a pseudorapidity of $|\eta| = 2.5$.  A homogeneous crystal electromagnetic
calorimeter (ECAL) and a sampling brass/scintilator hadronic calorimeter (HCAL)
surround the tracking volume and provide energy measurements of electrons, photons,
 and hadronic jets up to $|\eta| = 3.0$.
An iron-quartz fiber hadronic calorimeter, which is also part of the HCAL system,
is located in the forward
region defined by $3.0 < |\eta| < 5.0$.
Muons are measured in the pseudorapidity range of $|\eta| < 2.4$,
with detection planes made of three technologies: drift tubes, cathode strip chambers, and
resistive plate chambers. These are instrumented outside of the magnet coil within the steel
return yoke.   The CMS detector is nearly hermetic, allowing for energy balance measurements
in the plane transverse to the beam direction. A two-tier trigger system is designed to
select the most interesting $\Pp\Pp$ collision  events for use in physics analysis. A detailed
description of the CMS detector can be found elsewhere~\cite{CMS}.

\section{Reconstruction of Leptons, Missing Energy, and Jets}
\label{sec:objects}

Muon candidates are required to be successfully reconstructed~\cite{MUOPAS} using two algorithms, one
in which tracks in the silicon detector are matched to consistent
signals in the calorimeters and muon system, and another in which a
simultaneous global fit is performed to hits in the silicon tracker
and muon system.
The track associated with the muon candidate is required
to have a minimum number of hits in the silicon tracker,
have a high-quality global fit including a minimum number of hits in the muon
detectors, and have calorimeter energy deposits consistent with originating from a minimum ionizing particle.

Electron candidates are reconstructed~\cite{EGMPAS} starting from a
cluster of energy deposits in the
ECAL, which is then matched
to hits in the silicon tracker.
A selection using electron identification variables based on shower
shape and track-cluster matching is applied to the reconstructed
candidates; the criteria are optimized in the context of the inclusive
$\PW\rightarrow {\Pe\nu}$ measurement~\cite{inclusWXsect} and are
designed to maximally reject electron candidates from QCD multijet
production while maintaining approximately 80\% efficiency for electrons from the
decay of $\PW/\cPZ$ bosons.
Electron candidates within ${ \Delta
R}=\sqrt{{ \Delta\phi^2+\Delta\eta^2}}<$~0.1 of a muon are rejected to
remove electron candidates due to muon bremsstrahlung and final-state radiation.
Electron candidates originating from photon conversions are suppressed by looking for a partner track
and requiring no missing hits for the track fit in the inner layers of the tracking detectors.

Hadronic $\tau$ candidates ($\tau_h$) are identified~\cite{PFT-10-004}  starting with
a hadronic jet clustered from the particles
reconstructed using the particle-flow global-event reconstruction algorithm ~\cite{PFT-09-001}.
The highest-$p_T$\ charged track
within a cone of $\Delta\mathrm{R}< 0.1$ around the jet axis is required to have $p_T >$\ 5 GeV.
A variable size cone of $\Delta\mathrm{R} < 5$ GeV/$p_T$ is then defined around this track, and
the boosted $\tau$-decay products are expected to be confined within this narrow cone.
Only $\tau$ candidates with one or three charged hadrons in this cone
are selected. The discrimination between hadronic $\tau$ decays
and generic QCD jets is based on an ensemble of five neural networks, 
each of which has been trained to identify one of the five main hadronic $\tau$-decay modes
using the kinematics of the reconstructed charged and neutral pions~\cite{PFT-10-XXX}.

All lepton candidates are required to
have 
$|\eta|<$~2.4, and be consistent with originating from the same interaction vertex.
Charged leptons from the decay of $\mathrm{ W/Z}$ bosons, as well as the new physics
we are searching for, are expected to be
isolated from other activity in the event.
We calculate a relative measure of this isolation denoted as \RelIso.
This quantity is defined
as the ratio of the
scalar sum of transverse track momenta and transverse calorimeter energy deposits
within a cone of $\Delta R <$~0.3 around the lepton candidate direction at the origin,
to the transverse momentum of the candidate.
The contribution from
the candidate itself is excluded.

In order to suppress the background due to dileptons originating from the same jet,
we require that selected dileptons have a minimum invariant mass of 5 GeV. This
helps to keep dileptons uncorrelated with respect to their \RelIso observables, which is a feature we
exploit in the analysis. We also remove events with a third lepton of opposite sign and same
flavour as one of the two selected leptons if the invariant
mass of the pair is between 76 and 106 GeV.  This requirement
further reduces an already small background contribution
from WZ and ZZ production.

Jets
and \met are reconstructed based on the particle-flow technique desribed in~\cite{PFT-09-001,PFT-10-002}.  For jet clustering, we use the anti-$k_T$ algorithm
with the distance parameter $R = 0.5$~\cite{anti-kt}.
Jets are required to pass standard quality requirements~\cite{JME-10-001} to remove those consistent with calorimeter noise.
Jet energies are corrected for residual nonuniformity and nonlinearity of the detector response
derived using collision data~\cite{JES}.
We require jets to have transverse energy above 30~GeV and to be within $|\eta |<2.5$.
We define the $H_T$ observable as the scalar sum of the $p_T$ of all
such jets with ${ \Delta R > 0.4}$ to the nearest lepton passing all our requirements.

\section{Search Regions}
\label{sec:sr}

\begin{figure}[htp]
\begin{center}
\includegraphics[scale=0.25]{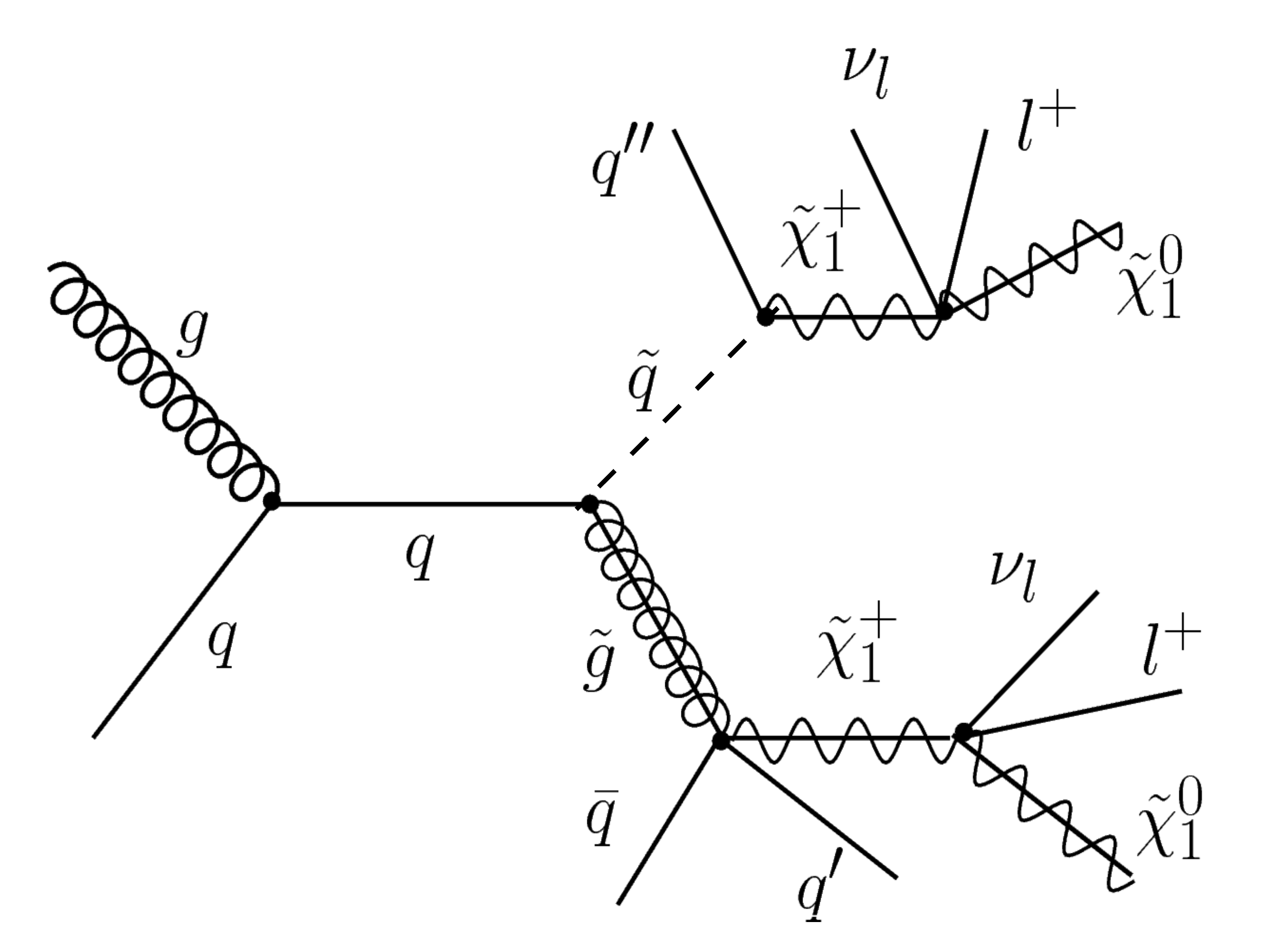}
\caption{An example of a process involving the production
and decays of SUSY particles, which  gives rise to two same-sign prompt leptons, jets, and missing transverse energy.}
\label{fig:FeynmanDiagram}
\end{center}
\end{figure}

The searches discussed in this paper
employ two different trigger strategies, electron and muon triggers in one case,
and $H_T$ triggers in the other.
The leptonic triggers allow for lower $H_T$ requirements, while the $H_T$ triggers
allow for lower lepton $p_T$, as well as final states with hadronic $\tau$ decays.
The motivation for covering the widest possible phase space in this search
can be illustrated by an example of a SUSY cascade, shown in Fig.~\ref{fig:FeynmanDiagram}, naturally giving
rise to jets, \met , and same-sign leptons:  (gluinos/squarks) $\rightarrow$
(charged gaugino) $\rightarrow$ (lightest supersymmetric particle (LSP) neutralino).  The mass difference between the gluino/squarks and the
charged gaugino,
typically arbitrary, defines the amount of hadronic activity one may
expect in the event. The mass difference between the gaugino and a
neutralino influences the lepton $p_T$ spectrum.
Depending on the nature of the
chargino and neutralino, their mass difference can be either arbitrary (e.g.,
wino and bino) or typically small (e.g., higgsinos). Moreover, there are
a number of ways to generate a large production asymmetry between $\tau$ and
$\Pe/\mu$ leptons, which motivates us to look specifically for events with
a $\tau$.

In the following we describe the search regions explored by each trigger strategy.
As a new-physics reference point, we use LM0, a point in the
constrained Minimal Supersymmetric Standard Model (CMSSM)~\cite{cmssm} defined with
the model parameters $m_0 = 200$\ GeV, $m_{1/2} = 160$\ GeV, tan$\beta = 10$, $\mu > 0$, and $A_0 = -400$ GeV.
LM0 is one of the common CMSSM reference points used in CMS across many analyses. As the abbreviation
suggests, LM0 provides squarks and gluinos with relatively low masses, and thus has a large production cross section.
 It is beyond the exclusion reach of the searches performed by LEP and Tevatron, but it  has
recently been excluded by searches~\cite{ATLAS1,SUS-10-003,SUS-10-007}
with ATLAS and CMS concurrently with this one.
Nonetheless, we continue to use LM0 as it provides a common model for which
to compare our sensitivity with that of other analyses.
Aside from the LM0 point, several SM simulation samples are used to both validate and complement various
background estimation methods that are based on the data itself.  These samples rely on either PYTHIA 6.4~\cite{pythia}
or MADGRAPH~\cite{madgraph} for
 event generation and GEANT4~\cite{GEANT4} for simulation of the CMS detector. Samples used
include $\ttbar$, single-t, $\gamma$+jets, W+jets, Z+jets, WW, WZ, ZZ, and QCD multijet production.
Next-to-leading-order (NLO) cross sections are used for all samples except for QCD multijet production.

\subsection{Searches using Lepton Triggers}
\label{sec:srlep}

We start  with a baseline selection inspired by our
published $\ttbar \to \ell^{+}\ell^{-} +\mathrm{X}$  ($\ell = \Pe$ or $\mu$)
cross section measurement~\cite{ref:top}.

Events are collected using single and dilepton triggers.
The detailed implementation of these triggers evolved throughout
the 2010 data-collecting period as the LHC instantaneous luminosity was increasing.
Trigger efficiencies are measured from a pure lepton sample
collected using $\cPZ \to \ell^{+} \ell^{-}$ decays from data.
The luminosity-averaged
efficiency to trigger on events with two leptons
with $|\eta | < 2.4$ and
$p_T >$ 10~GeV, one of which also has $p_T > $ 20~GeV, is very high.
For example, the trigger efficiency for an LM0 event passing the
baseline selection described below is estimated to
be $(99 \pm 1)$\%.

One of
the electrons and muons
must have $p_T > 20$ GeV and the second one must have $p_T > 10$~GeV.
Both leptons must be isolated.
The isolation requirement is based on the \RelIso variable introduced earlier.
We require \RelIso $< 0.1$ for leptons of $p_T > 20$ GeV, and
the isolation sum (i.e., the numerator of the \RelIso expression) to be less than 2 GeV for $p_T < 20$ GeV.

We require the presence of at least two reconstructed jets, implying $H_T > $ 60 GeV.
Finally, we require the missing transverse energy
\met $> 30$ GeV ($\Pe\Pe$ and $\mu\mu$) or \met $ > 20$ GeV ($\Pe\mu$).
This defines our {\it baseline selection}.

Following the guiding principles discussed in the introduction, we define two search regions.
The first 
has high \met\ (\met $> 80$ GeV);
the second 
has high $H_T$ ($H_T >$ 200 GeV).
These \met\ and $H_T$ values were chosen to obtain an
SM background expectation in simulation
of 1/3 of an event in either of the two overlapping search regions.

\begin{figure}[bht]
\begin{center}
\includegraphics[width=0.48\linewidth]{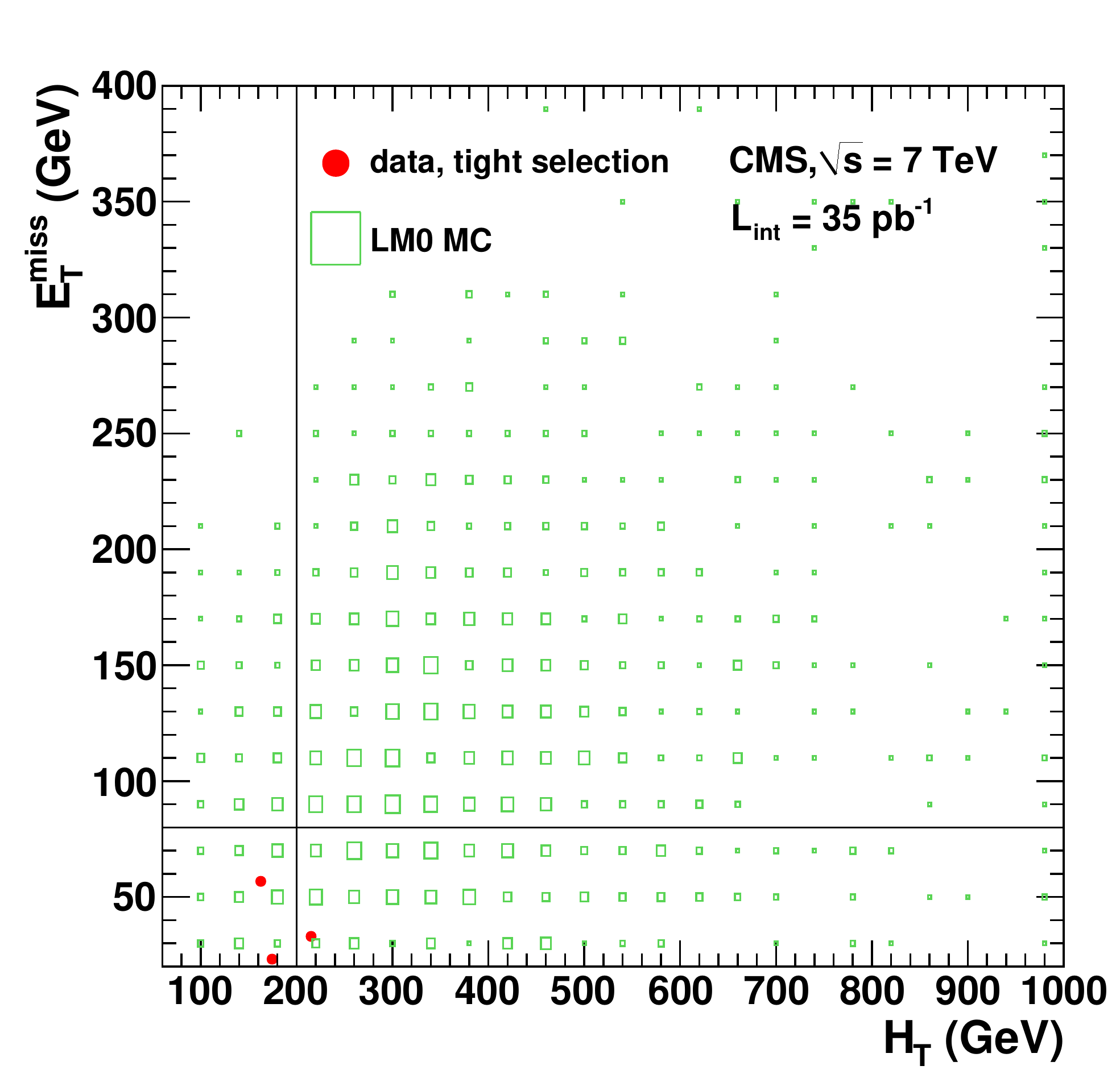}
\includegraphics[width=0.48\linewidth]{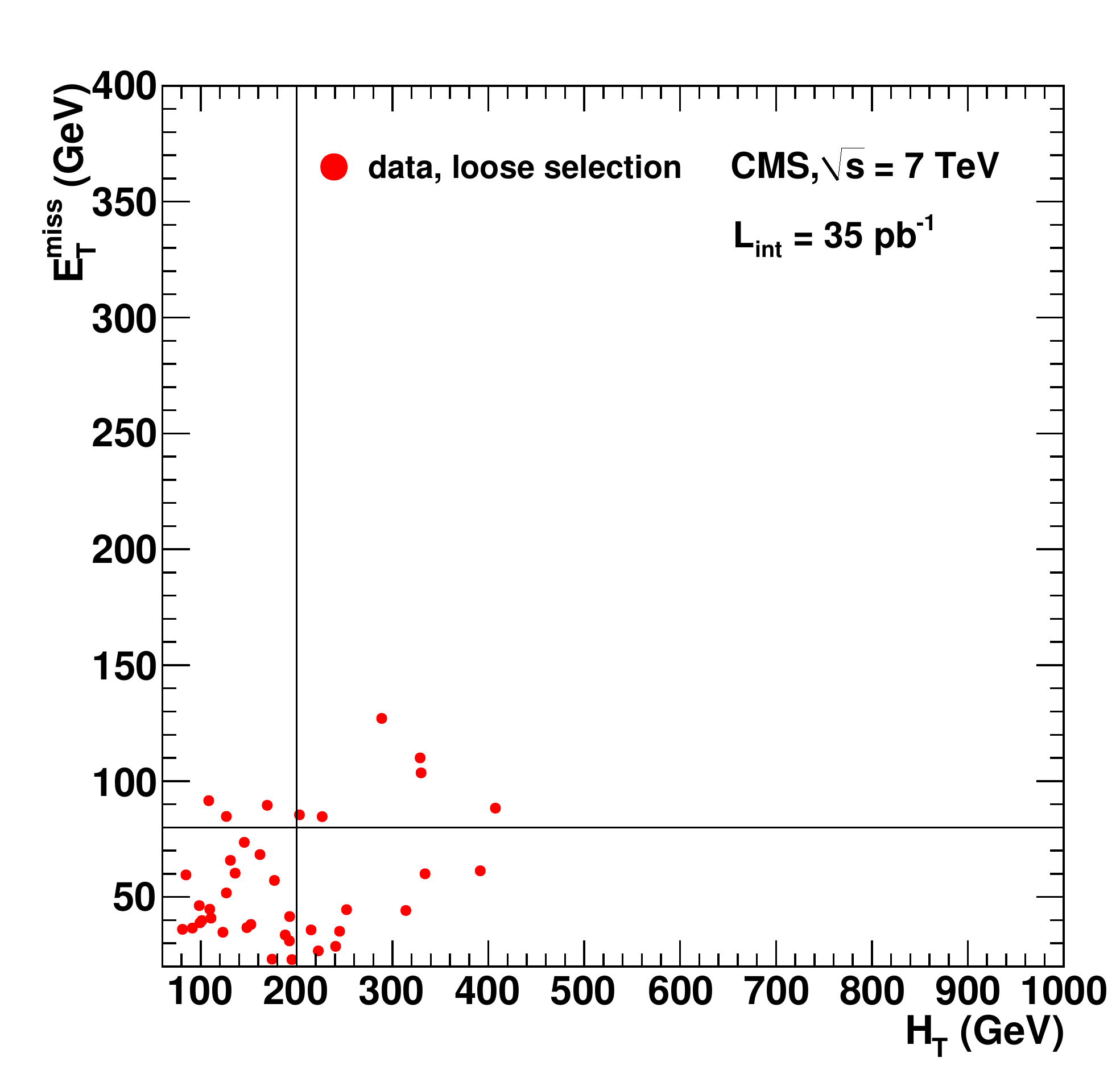}
\caption{\label{fig:metvsht}\protect
$H_T$ versus
\met\ scatter plots for baseline region.
(Left) Overlay of the three observed events with
the expected signal distribution for LM0. The three observed events all scatter in the lower left corner of the plot.
(Right) Scatter plot of the background in data when only one of the
two leptons is required to be isolated.
}
\end{center}
\end{figure}

Figure~\ref{fig:metvsht} shows the $H_T$ versus \met\ scatter plot for the baseline selection, indicating
the \met\ and $H_T$ requirements for the two search regions via horizontal and vertical lines, respectively.
Figure~\ref{fig:metvsht} (left) shows three events (red dots) in the baseline region,
one of which barely satisfies $H_T > $ 200 GeV,
but fails the \met\ $>$ 80 GeV requirement.
In contrast, most of the signal from typical supersymmetry models
tends to pass both of these requirements, as is visible in the LM0 expected signal distribution overlaid in Fig.~\ref{fig:metvsht} (left).
Backgrounds to this analysis are dominated by events with jets mimicking leptons, as discussed in Section~\ref{sec:bkg}.
Requiring only one of the two leptons to be isolated thus allows us to increase the background statistics in order to
display the expected distribution of SM background events in the $($\met$,H_T)$ plane,
as shown in Fig.~\ref{fig:metvsht} (right).
Backgrounds clearly cluster at low \met and low $H_T$, with slightly more than half of the events failing both the \met\ and $H_T$ selections.
Moreover, comparing the left and right plots in Fig.~\ref{fig:metvsht}
indicates that the lepton isolation requirement on both leptons versus only one lepton reduces the
backgrounds by roughly a factor of ten.

\subsection{Searches using Hadronic Triggers}
\label{sec:srhadtrig}

\begin{figure}[b]
\begin{center}
  \begin{minipage}{0.5\columnwidth}
  \resizebox{8cm}{!} {\includegraphics{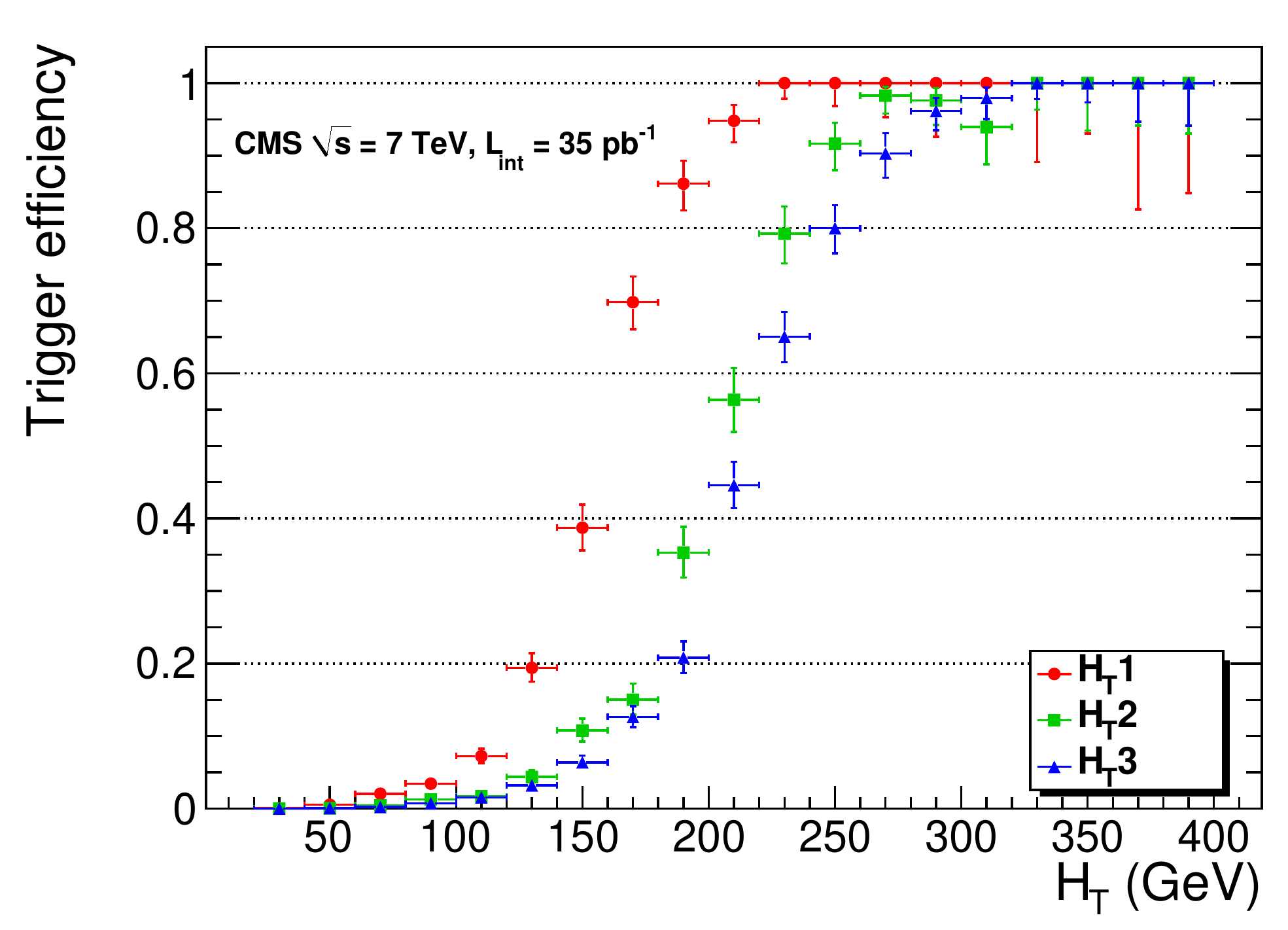}}
  \end{minipage}
  \hfill
\caption{$H_T$ Trigger efficiency as a function of the reconstructed $H_T$ for three data-collecting periods:
7 pb$^{-1}$ with $H_{T}1$, 10 pb$^{-1}$ with $H_{T}2$, and 18 pb$^{-1}$ with $H_{T}3$.
}
\label{fig:SignalEff}
\end{center}
\end{figure}

Hadronic triggers allow us to explore the phase space with low-$p_T$ electrons and muons,
as well as final states with hadronic $\tau$ decays.
We allow muons (electrons) with $p_T$ as low as 5 (10) GeV, and
restrict ourselves to $\tau_h$ with visible transverse momentum $> 15$~GeV,
where $\tau_h$ refers to hadronic $\tau$ candidates only.
All leptons must be isolated with \RelIso $< 0.15$.

For the $\Pe\Pe, \Pe\mu$, and $\mu\mu$ final states, we
require at least two jets, $H_T > 300$ GeV,
and \met $> 30$ GeV.
As backgrounds from QCD multijet production are significant for $\tau_h$, we increase the \met\ and $H_T$ requirements to \met\ $> 50$\ GeV and $H_T > 350$\ GeV
in the $\Pe\tau_h$, $\mu\tau_h$, and $\tau_h\tau_h$ final states.

Figure~\ref{fig:SignalEff} shows the efficiency turn-on curves for the $H_T$ triggers used during three different data taking
periods. The trigger thresholds were changing in order to cope with the increasing instantaneous luminosities over the 2010 running
period.  Roughly half of the integrated luminosity in 2010 was taken with the highest threshold trigger.
This measurement indicates that at $H_T = $ 300 GeV the efficiency reaches $(94\pm 5)$\% .
These trigger turn-on curves are measured in data with events selected by muon triggers.

\section{Background Estimation}
\label{sec:bkg}

Standard model sources of same-sign dilepton events with both leptons coming from a $\PW$ or $\cPZ$ decay
are very small in our data sample.
Simulation-based predictions of the combined yields for
$\Pq\Paq \rightarrow \PW\cPZ$ and ZZ, double ``W-strahlung'' $\mathrm{qq \rightarrow q'q'W^{\pm}W^{\pm}}$,
double parton scattering $2 \times (\Pq\Paq \rightarrow \PW^{\pm})$, $\ttbar\PW$, and WWW
comprise no more than a few percent of the total background in any of the final states considered.
As these processes have never been measured in proton-proton collisions,
and their background contributions are very small,
we evaluate them using simulation,
assigning a 50\% systematic uncertainty.
The background contribution from $\Pp\Pp \to \PW \gamma$, where the W decays leptonically
and the photon converts in the detector material giving rise to an isolated
electron, is also estimated from simulation and found to be negligible.
All other backgrounds are evaluated from data, as discussed below.

Backgrounds in all of our searches are dominated by one or two jets mimicking
the lepton signature. Such lepton candidates can be genuine leptons from heavy-flavour decays,
electrons from unidentified photon conversions, muons from meson decays in flight,
hadrons reconstructed
as leptons, or jet fluctuations leading to hadronic $\tau$ signatures.
We will refer to all of these as "fake leptons".
Leptons from W, Z, gauginos, etc., i.e., the signal we are searching for,
will be referred to as "prompt leptons".

The dominant background contribution is from events with one lepton, jets, and \met---mostly $\mathrm{t\bar{t}}$
with one lepton from the W decay, and a second lepton from the decay of
a heavy-flavour particle.
These events contain one prompt and one fake lepton, and are estimated via two different techniques described in
Sections~\ref{sec:leptrg} and~\ref{sec:Florida}.
While both techniques implement an extrapolation in lepton isolation, they differ in the assumptions made.
Both techniques lead to consistent predictions
as described in Section~\ref{sec:compare}, providing additional confidence in the results.
Backgrounds with two fake leptons are generally smaller,
except in the final state with two hadronic $\tau$ leptons, where the dominant background source is
QCD multijet production.
Contributions due to fake $\tau_h$ are estimated using an extrapolation
from ``loose'' to ``tight'' $\tau_h$ identification, as described in Section~\ref{sec:taufake}.

For the $\Pe\Pe$ and $\Pe\mu$ final states, electron charge misreconstruction due to hard bremsstrahlung poses
another potentially important
background, as there are significant opposite-sign $\Pe\Pe$ and $\Pe\mu$ contributions, especially from
$\mathrm{t\bar{t}}$, where both W's from the top quarks decay leptonically.
This is discussed in Section~\ref{sec:flip}.

\subsection{Searches using Lepton Triggers}
\label{sec:leptrg}

Contributions from
fake leptons
are estimated using the so-called
``tight-loose'' (TL) method~\cite{ref:top,ref:FR}.
In this method the probability $\epsilon_{TL}$
for a lepton passing loose selections to also pass the tight analysis
selections is measured in QCD multijet events as a function of lepton
$p_T$ and $\eta$.
The key assumption
of the method is that $\epsilon_{TL}$ is approximately universal, {i.e.},
it is the same for all jets in all event samples.
Tests of the validity of this assumption are described below.

The main difference between the tight and loose
lepton selections is that the requirement on the \RelIso variable defined in
Section~\ref{sec:objects} is relaxed from \RelIso $< 0.1$ to
\RelIso $< 0.4$. Other requirements that are relaxed are those
on the distance
of closest approach between the lepton track and the beamline (impact parameter)
and, in the case of muons, the selection on the
$\chi^2$ of the muon track fit.

The quantity $\epsilon_{TL}$ is measured
in a sample of lepton-trigger events with at least one jet satisfying $p_T > 40$ GeV
and well separated ($\Delta R > 1$) from the lepton candidate.
We refer to this jet as the ``away-jet''.
We reduce the impact of electroweak background
(W, Z, $\ttbar$) by excluding events with
$\cPZ \to \ell \ell$ candidates, events with \met\ $> 20$ GeV,
and events where the transverse mass $M_T$ of the lepton and the \met\
is greater than $25$ GeV.  Studies based on simulation indicate that this procedure
results in an unbiased estimate of $\epsilon_{TL}$ up to
lepton $p_T \approx 40$ GeV.  At higher transverse lepton momenta the
remaining electroweak contributions in the sample have
a significant effect.
Thus, $\epsilon_{TL}$ is measured only up to $p_T = 35$ GeV.
It is taken to be constant at higher transverse momenta, as
suggested by simulation studies.

\begin{figure}[tbh]
\begin{center}
\includegraphics[width=0.48\linewidth]{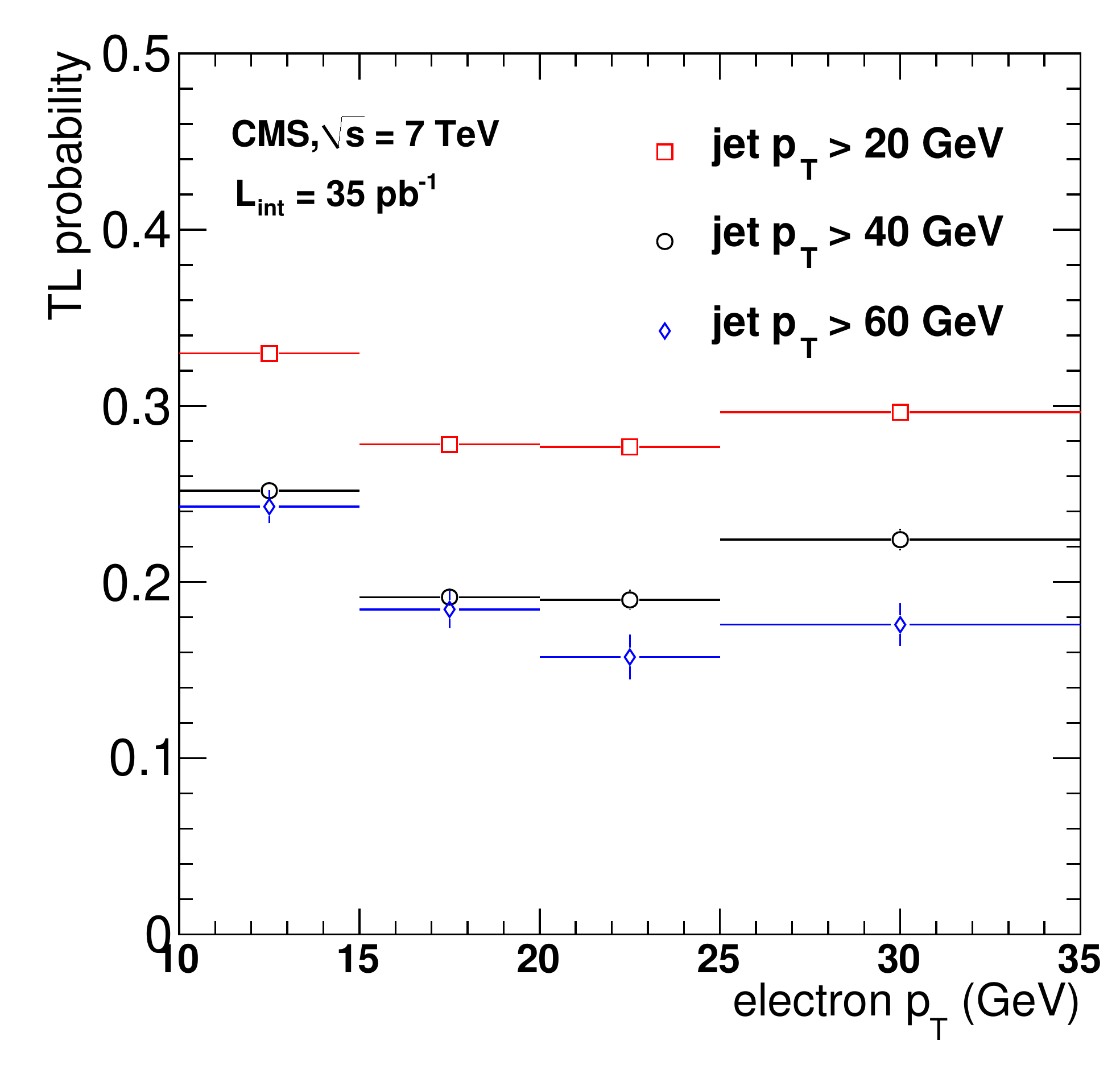}
\includegraphics[width=0.48\linewidth]{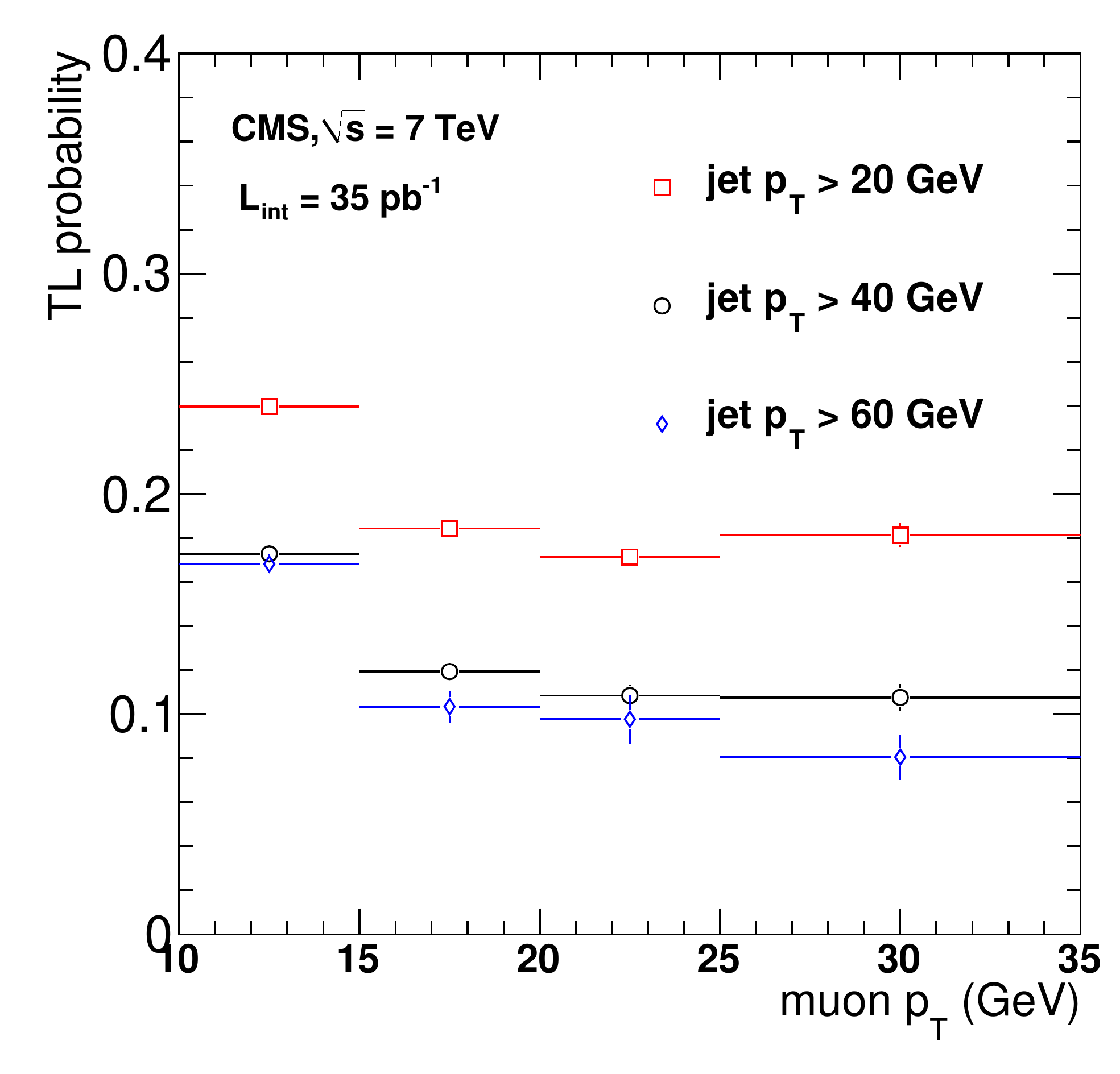}
\caption{\label{fig:FRjet}\protect Electron (left) and muon (right)
TL probability $\epsilon_{TL}$
computed from QCD multijet events with different requirements
on the minimum $p_T$ of the away-jet. The probabilities shown are projections of the two-dimensional function
$\epsilon_{TL}(\eta , p_T )$ onto the $p_T$ axis.}
\end{center}
\end{figure}

The level of universality of $\epsilon_{TL}$ is tested with different
jet samples.  
Two types of tests are relevant and both
involve the parent jet from which the lepton originates. The first
test explores the sensitivity to the jet's $p_{T}$, and the second test
explores the sensitivity to the jet's heavy-flavour content.

Sensitivity to jet $p_T$ stems from the fact
that the probability for a lepton of a given $p_T$
to pass the \RelIso selection depends on the $p_T$ of the
parton from which the lepton originates.
To be explicit, a 10~GeV lepton originating from a 60 GeV b quark is less likely to pass our \RelIso requirement than
the same lepton originating from a 20 GeV b quark.
The heavy
flavour sensitivity can be traced to
semileptonic decays, which are a source of leptons in bottom
and charm jets, but not in light-quark and gluon jets.

To test the jet $p_T$ dependence, we select loose leptons
in events with the away-jet above a varying jet $p_T$ threshold.
Since these events are mostly QCD dijets, the $p_T$ of
the away-jet is a good measure of the $p_T$ of the jet
from which the
lepton originates.  We weight each event by
$\epsilon_{TL}$ measured as described above, {i.e.},
requiring the away-jet to have $p_T > 40$ GeV.  We then
sum the weights and compare the sum to the
number of observed leptons passing tight requirements.
Varying the away-jet minimum $p_T$ requirement from 20 to 60 GeV, we find the
observed yield to differ from the predicted yield by +54\% (+49\%) and $-$4\% ($-$3\%)
for muons (electrons) in this test. The percentages here, as well as throughout this section refer to
(observed~$-$~predicted)/predicted.
This non-negligible jet $p_T$ dependence can also be
seen in Fig.~\ref{fig:FRjet}, where we show $\epsilon_{TL}$
calculated using different away-jet thresholds.
To test the heavy-flavour dependence,
we repeat the exercise
requiring that the
away-jet be above $p_T > 40$ GeV and be b-tagged, {i.e.},
a jet in which we find a secondary vertex well separated
from the interaction point consistent with a b-hadron decay.
By applying the b tag on the away-jet, the sample
of jets from which the lepton originates is enriched in heavy
flavours.
Introducing this b tag we find the observed yield to differ from that predicted by $-$3\% ($-$15\%) for muons (electrons).
We have thus shown that
applying an $\epsilon_{TL}$ obtained without a b-tagging requirement to a sample with such a requirement leads to a modest
difference between observed and predicted. This validates our assumption that the TL method is flavour universal.

To predict the background from prompt lepton $+$ jets events in a signal region,
the TL probability is applied
to a sample of dilepton events satisfying all the signal selection requirements, but
where one of the leptons fails the tight selections and passes the loose ones.
Each event is weighted by the factor $\epsilon_{TL}/(1-\epsilon_{TL})$,
where $\epsilon_{TL}$ is the tight-to-loose probability for the
loose lepton in the event.  The background contribution
from this source is then estimated
by summing the weights of all such events
(${\cal S}_1$).

The sum ${\cal S}_1$
also includes the contribution from backgrounds with
two fake leptons.  However, these are double counted
because in the case of two fake leptons passing
loose requirements there are two combinations with
one lepton passing the tight selections.  The background
contribution with two fake leptons is estimated separately by selecting
events where both leptons pass the loose requirements but fail
the tight requirements.  Each event in this sample is weighted
by the product of the two factors of
$\epsilon_{TL}/(1-\epsilon_{TL})$ corresponding to the
two leptons in the event, and the sum ${\cal S}_2$ of weights
is used to estimate the background with two fake leptons.

The total background from events with one or two fake leptons
is then obtained as ${\cal S}_1 - {\cal S}_2$.
In kinematic regions of interest for this search,
${\cal S}_2$ is typically more than one order of magnitude
smaller than ${\cal S}_1$, indicating that the
main background contribution is from one prompt lepton and
one fake lepton.

The method has been tested on simulated $\mathrm{t\bar{t}}$ and
W $+$ jets events.  In these tests, we
use $\epsilon_{TL}$ measured from QCD simulation events
to predict the number of same-sign dilepton events
in these samples.   In the $\mathrm{t\bar{t}}$ simulation
sample, we find that the
observed yield differs from the prediction
for the baseline selection by -41\% (fake muons) and -47\%
(fake electrons).
Observed and predicted yields in this test are consistent with each other at the 5\% confidence level (CL) because
the simulation statistics are modest.
The same level of agreement is also found for the two search regions.
In the W $+$ jets case, the ratio of predicted events to observed is $0.8\pm 0.4$ for
fake electrons; the statistics for fake muons are not
sufficient to draw any definitive conclusions.   Based on these
studies, as well as the dependence on away-jet $p_T$ and heavy-flavour composition discussed above,
we assign a $\pm 50\%$ systematic uncertainty on the ratio (observed~$-$~predicted)/predicted
and, hence, on the estimation of backgrounds due to fake leptons.
In addition to this systematic uncertainty, the method has significant statistical uncertainties based on the number of events
in the samples to which the TL probability is applied. We find $6$\ $(4)$ events in these samples for the
\met $> 80$ GeV
($H_T > 200$ GeV) search regions.
The resulting background estimates in the two regions are
$1.1\pm 0.6$ and $0.9\pm 0.6$ events, respectively, including only statistical uncertainties.

As an additional cross-check, we determine the background estimate and observed yields in the baseline region.
We estimate $3.2\pm 0.9\pm 1.6$ events from background due to fake leptons alone, and $3.4\pm 0.9\pm 1.6$ after all backgrounds are taken into account.
The uncertainties here are statistical and systematic, respectively.
The composition of the total background is estimated to be 86\% (7\%) events with one (two) fake leptons, 3\% due to charge misidentification,
and 4\% irreducible background for which both leptons are isolated leptons from leptonic W or Z decay.
As mentioned in Section~\ref{sec:srlep}, there are 3 events observed
in the baseline region, in good agreement with the background estimate.
Applying \met $>$ 80 GeV or $H_T >$ 200 GeV increases the fraction of events with one fake lepton, but statistical
uncertainties on the individual components of the background estimate are too large to meaningfully quantify
the change in the relative contributions
of the components.

\subsection{Search using Hadronic Triggers, Electrons, and Muons}
\label{sec:Florida}

As described in Section~\ref{sec:srhadtrig}, hadronic triggers allow us to explore the phase space with low-$p_T$ leptons.
However, lowering the lepton $p_T$ is expected to increase the relative contribution of events with two fake leptons.
As shown below,
this background
now constitutes roughly 30\% of the total background, as compared to only
 a few percent for the higher-$p_T$ thresholds in the search regions for the leptonic trigger analysis.
This motivates
the development of a method exclusively dedicated to predicting and understanding the QCD background
with two fake leptons.

At the same time, increasing the $H_T$ requirement to 300 GeV, as driven by the hadronic trigger thresholds,
reduces the expected W+jets background to only a few percent of the total background.
The background with one fake lepton is now reduced to $\ttbar$ and single-t processes, where
the fake lepton is due mostly to semileptonic b decays.
Therefore, we tailor the method for estimating
the background with one fake lepton to these expectations.
The method is similar to the TL technique, but has a number of important differences that are discussed further below.

For this analysis,
the estimation of the background with fake leptons starts with an evaluation of background events with
 two fake leptons and then proceeds with an estimation of the contribution of events with only one such lepton.

First, we define a \textit{preselection} control sample of events from the $H_T$-triggered data stream with at least
two same-sign dileptons and with all event selection requirements applied, except for those related to \met and isolation.
We find 223 $\mu\mu$, 6 $\Pe\Pe$, and 78 $\Pe\mu$ events of this type.
The large asymmetry between muons and electrons is mostly due to differences in the
corresponding $p_T$ thresholds of 5 and 10 GeV, respectively.
In addition, identification of electrons within jets is less efficient than that for muons.

The preselection control sample is dominated by QCD multijet production.
Studies based on simulation suggest that we should attribute about 10\% of
the preselection yields to $\ttbar$ contamination, while attributing a much smaller fraction to W+jets.

The contribution from events with two fake leptons to the signal region is estimated by assuming that the three requirements,
\RelIso $< 0.15$ for each lepton and \met $>$ 30 GeV, are mutually independent and, hence, the total
background-suppression efficiency can be written
in the factorized form
$\epsilon_{\mathrm{tot}} = \epsilon_{\ell_{1} \, \mathrm{iso}} \cdot \epsilon_{\ell_{2} \, \mathrm{iso}} \cdot \epsilon_{\mathrm{MET}}$.
This assumption has been verified both in simulation and directly in data. With simulation it is straightforward to
prove the principle in the nominal {preselection} region because we can safely measure the efficiencies in a
dedicated QCD sample where we know all leptons can be considered background (i.e., no contamination from prompt leptons exists).
 In data the contribution from prompt leptons is non-negligible and therefore some extra selection requirements
are necessary to isolate a QCD enriched control sample.

\begin{figure}[h]
  \begin{minipage}{0.5\columnwidth}
  \resizebox{7cm}{!} {\includegraphics{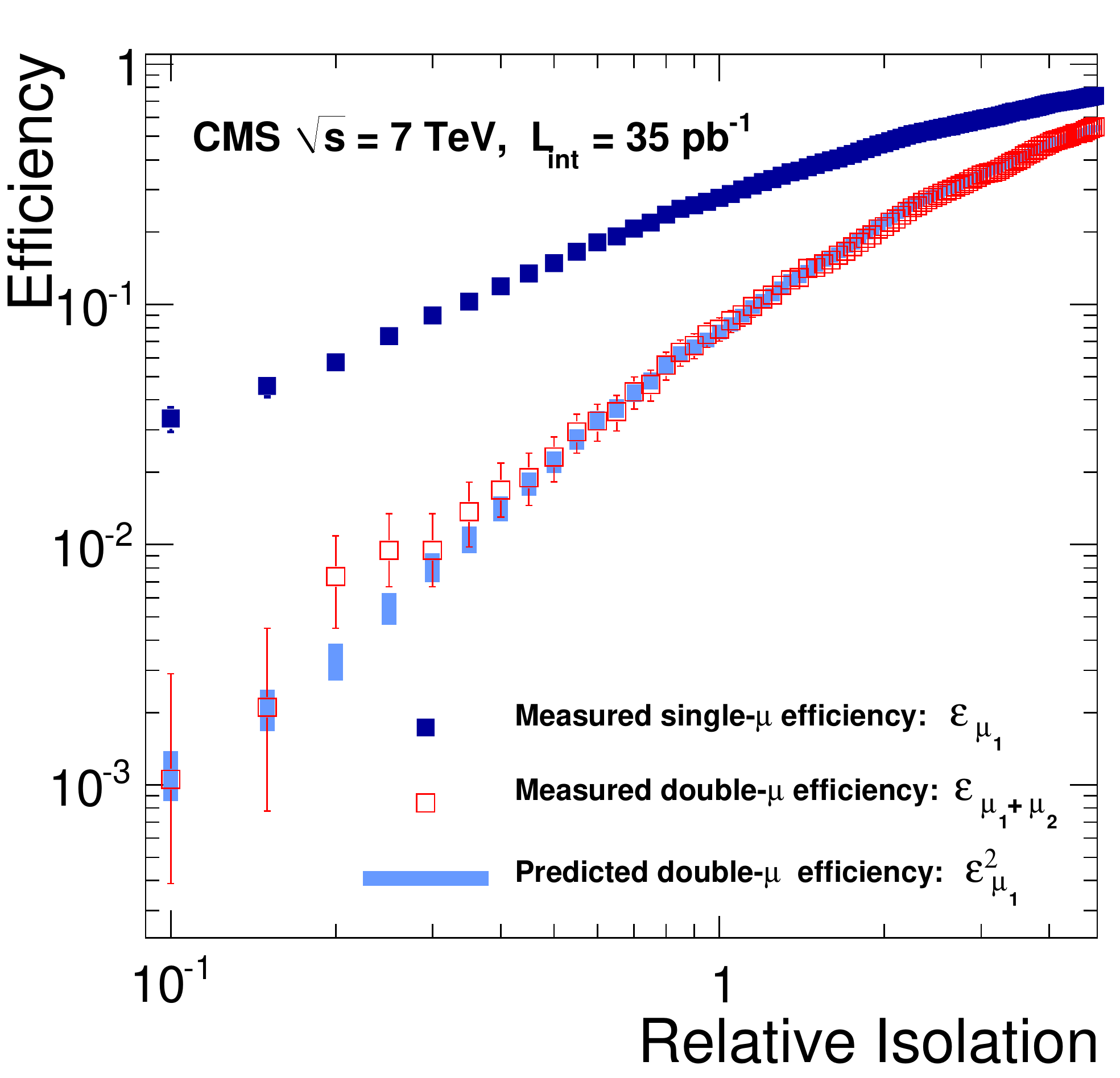}}
  \end{minipage}
  \hfill
  \begin{minipage}{0.5\columnwidth}
  \resizebox{7cm}{!} {\includegraphics{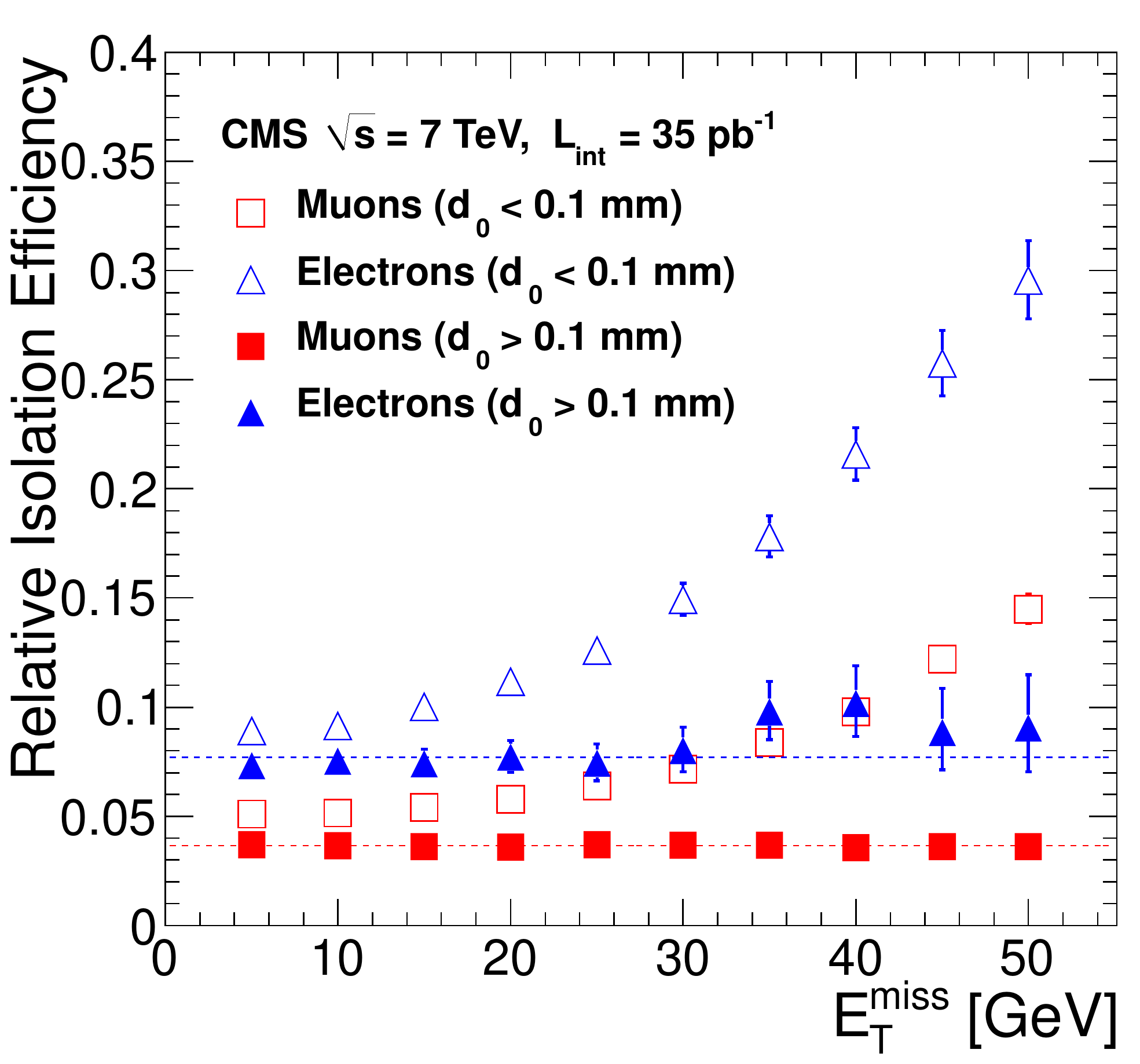}}
  \end{minipage}
\caption{
(Left)
The lepton isolation efficiency for one (solid squares) and two (open squares) leptons as a function of the
relative isolation parameter cut.
Also shown is the predicted double-lepton efficiency if the two lepton efficiencies are assumed to be independent of each other.
Only the dimuon sample is shown here.
(Right)
The lepton isolation efficiency as a function of the \met cut for electrons and muons with different requirements on the
lepton impact parameter.
Details are given in the text.
}
\label{fig:uncorrelated}
\end{figure}

We validate the factorized expression for $\epsilon_{\mathrm{tot}}$ in two steps
in data. First, we demonstrate that the selection requirement on \RelIso is independent for each lepton.  We begin by relaxing
the $H_T$ selection to 200 GeV and add events collected with leptonic triggers to gain more statistics.
We then require \met $<$ 20 GeV to suppress events with leptonic W decays. Figure~\ref{fig:uncorrelated} (left)
shows that the  single-muon efficiency can be squared to obtain the double-muon efficiency, thus validating the
assumption that the \RelIso observable is uncorrelated between the two fake leptons and the efficiencies can be factorized.
In the second step, we demonstrate that the \met and \RelIso selection requirements are mutually independent. To
accomplish this in data, we maintain $H_T$ above 300 GeV, but we include single-lepton events to increase statistics.
To suppress the contributions from events with leptonic W decays, we modify the selection requirement on the
lepton impact parameter from the nominal $d_0 < 0.2$ mm to  $d_0 > $ 0.1 mm.  Figure~\ref{fig:uncorrelated} (right) shows that the \RelIso
selection efficiency for muons and electrons remains constant as a function of the \met selection
requirement. The dashed lines represent the zeroth-order polynomial fits to the efficiency measurements made
 in the $d_{0}$ control region for muons and electrons, respectively.  For completeness, we also show the
obvious bias arising when the impact parameter requirement
is inverted to $d_0 < $ 0.1 mm to enrich the sample with leptonic W decays.
It is important to note that no attempt is made to apply the \RelIso selection efficiency measured in the
control region defined by $d_0 > $ 0.1 mm to the above formula for $\epsilon_{\mathrm{tot}}$.
This control region is only used to demonstrate the stability of the  \RelIso selection efficiency
with respect to the \met requirement.  The actual values of $\epsilon_{\ell_{1} \, \mathrm{iso}}$ and
 $\epsilon_{\ell_{2} \, \mathrm{iso}}$ are measured in the nominal {preselection} region ($d_0 < $ 0.2 mm), where
we assume this stability, and hence factorization, remains valid for events with two fake leptons.

Having validated the selection factorization hypothesis, we proceed to measure the isolation and \met selection
efficiencies, one at a time, in the {preselection} control sample, where we obtain
$\epsilon_{\mu \, \mathrm{iso}} = 0.036\pm 0.015$,
$\epsilon_{\Pe \, \mathrm{iso}} = 0.11\pm 0.08$,
$\epsilon_{\mathrm{MET}} = 0.27\pm 0.03$.
Uncertainties quoted are statistical only.
As before, we suppress leptonic W decays to reduce possible biases.
We accomplish this by requiring either \met $<$ 20 GeV or \RelIso $>$ 0.2
when measuring $\epsilon_{\mu \, \mathrm{iso}}$, $\epsilon_{\Pe \, \mathrm{iso}}$, or $\epsilon_{\mathrm{MET}}$.
The appropriate product of these efficiencies is then applied to
the event counts observed in the {preselection} control sample,
leading to the background estimate of $0.18 \pm 0.12 \pm 0.12$ events.
Uncertainties here are statistical and systematic, respectively.

The systematic uncertainties quoted above have two dominant sources.
One is due to limited statistics in simulation and data when validating that the three requirements are
indeed independent. We take the statistical precision (25\%) of this cross-check as our systematic
uncertainty on the method. The other dominant source of systematic uncertainty can be attributed to the inability
of the inverted selection requirements on \met and \RelIso to fully suppress contributions from leptonic W decays
(e.g., $\ttbar$, W+jets)
while measuring  $\epsilon_{\mu \, \mathrm{iso}}$, $\epsilon_{\Pe \, \mathrm{iso}}$, and $\epsilon_{\mathrm{MET}}$.
Studies based on simulation suggest that the bias (overestimate) can be as large as 60\%, mostly via a
bias in measuring $\epsilon_{e\ iso}$. Conservatively, we do not correct for the possible bias, but
take it as a systematic uncertainty. We thus arrive at a 65\% systematic
uncertainty on the estimate of backgrounds due to
events with two fake leptons by adding these two effects in quadrature.
In addition, we verified in simulation that the techniques used to suppress leptonic W decays
(i.e., inversion of the \met and \RelIso requirements) do not alter or bias the selection efficiencies
for fake leptons.

It is worth mentioning that this method of evaluating background with two fake leptons does not require
any reweighting of measured efficiencies. The average efficiencies are obtained from a QCD-dominated subset of the
{preselection} sample, and then applied directly to the {preselection} sample as a whole to derive the
prediction for the number of events with two fake leptons in the signal region.

Next, we proceed with estimating the contribution of backgrounds with a single misidentified lepton.
We start from a tight-loose control sample, to be further referred to as a \textit{sideband},
and use the isolation selection efficiency
for b jets, referred to as $\epsilon^{(\cPqb)}$, to predict event counts in the signal region.
The {sideband} control sample contains events passing all signal selection criteria,
except one of the two leptons is now required to have \RelIso $>$ 0.15.
To begin, we count the number of events in this sample: 11 ($\mu\mu$), 2 ($\Pe\Pe$), 6 ($\Pe\mu$), and 5 ($\mu\Pe$),
the last lepton indicating which one in the pair is non-isolated.
Then, we estimate the contribution of the background with two fake leptons to the {sideband} sample
using the efficiencies quoted above from the factorization procedure.
For example, for the dimuon channel, the contribution to the sideband from events with two fake leptons
is $N_{\mu\mu \, \mathrm{preselected}} \cdot 2 \epsilon_{\mu \, \mathrm{iso}} (1 - \epsilon_{\mu \, \mathrm{iso}}) \cdot \epsilon_{\mathrm{MET}}$.
The resulting yield estimates for events with two fake leptons are
4.2 ($\mu\mu$),
0.32 ($\Pe\Pe$),
2.3 ($\Pe\mu$), and
0.68 ($\mu\Pe$).
After subtracting this contribution, the remaining yields in the sideband are consistent with simulation predictions
assuming that only $\ttbar$ (76\%), single-t (7\%), and W+jets (15\%) contribute.
This remaining sideband yield after subtraction is then scaled by an appropriate factor determined
using the \textit{BTag-and-probe} method~\cite{ref:FR}, as described below.

The {BTag-and-probe} method relies on the basic premise that events with one fake lepton
can be attributed to $\ttbar$ production, with one prompt lepton from  leptonic W decay and
the second fake lepton from semi-leptonic b decay.
The efficiencies $\epsilon_{\mu \, \mathrm{iso}}^{(\cPqb)}$ and $\epsilon_{\Pe \, \mathrm{iso}}^{(\cPqb)}$ are thus defined as the probabilities
of a muon or electron from semi-leptonic $\cPqb$ decay to pass the \RelIso $<$ 0.15 selection.
These efficiencies can be measured in data using appropriately selected events from $\bbbar$ production.
To determine $\epsilon_{\mu \, \mathrm{iso}}^{(\cPqb)}$ and $\epsilon_{e \, \mathrm{iso}}^{(\cPqb)}$, we select a $\bbbar$
enriched
control sample by requiring one b-tagged away-jet and one lepton candidate.
In addition, we require $H_T > 100$ GeV to arrive at a b-quark $p_T$ spectrum similar to that expected for
the $\ttbar$ background in the search.  To reduce the bias from leptonic W or Z decays, we furthermore require
\met $<$ 15 GeV and $M_T < $ 15 GeV, and veto events with two leptons
forming a mass within 7 GeV of the mass of the Z boson.
Approximately 80\% of the leptons in this sample are from semileptonic heavy-flavour decay.

\begin{figure}[h]
  \begin{minipage}{0.5\columnwidth}
  \resizebox{8cm}{!} {\includegraphics{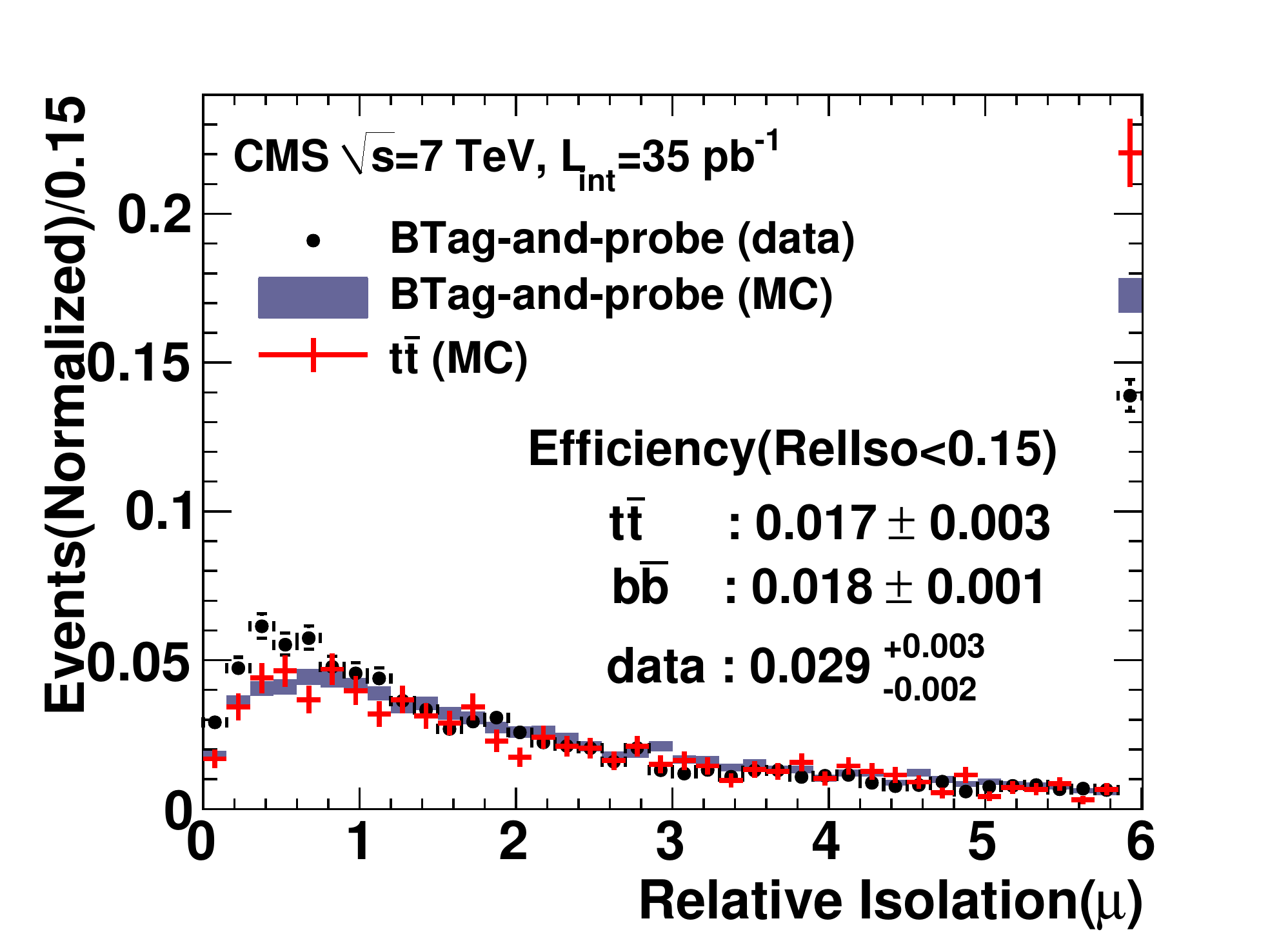}}
  \end{minipage}
  \hfill
  \begin{minipage}{0.5\columnwidth}
  \resizebox{8cm}{!} {\includegraphics{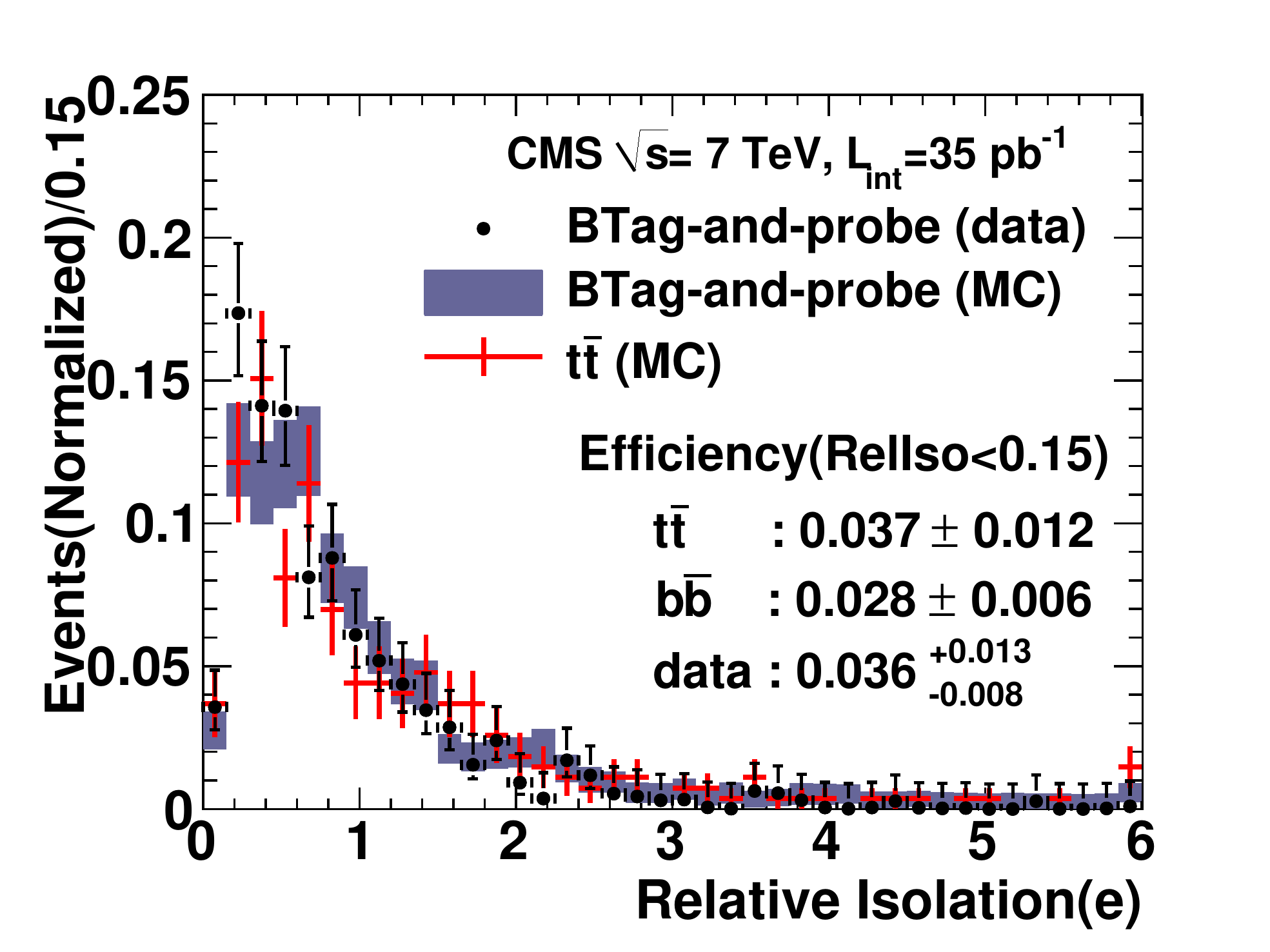}}
  \end{minipage}
\caption{Isolation variable distributions obtained with the {BTag-and-probe} method for muons (left) and electrons (right).
Efficiencies for the \RelIso $<$ 0.15 (first bin in the distributions shown) are explicitly quoted.}
\label{fig:ttbar}
\end{figure}

We find that the resulting $\mathrm{b\bar{b}}$ control sample differs sufficiently from the expected
$\ttbar$ background
in both lepton kinematics and jet multiplicity ($N_{\mathrm{jets}}$) to warrant corrections.
We therefore measure the \RelIso distribution in the $\bbbar$ control sample in data in bins of lepton $p_T$ and $N_{\mathrm{jets}}$, and
reweight these distributions using event probabilities $\omega(p_T, N_{\mathrm{jets}})$ derived from a
$\mathrm{t\bar{t}}$ simulation sample.
The resulting reweighted \RelIso distributions for these three samples are overlaid in
Fig.~\ref{fig:ttbar} for muons (left) and electrons (right).
The plots show distributions for $\mathrm{t\bar{t}}$ simulation (red crosses) after all selections except \RelIso on one of the two leptons,
reweighted $\mathrm{b\bar{b}}$ simulation (grey shade), and reweighted $\mathrm{b\bar{b}}$-enriched data (black dots).
The agreement between the two simulation-based distributions validates the method.
Agreement between data and simulation is observed but is not required for this method to be valid.
The contents of the first bin of the two data plots are the above mentioned isolation selection efficiencies
$\epsilon_{\mu \, \mathrm{iso}}^{(\cPqb)}$ and $\epsilon_{\Pe \, \mathrm{iso}}^{(\cPqb)}$.
We find
$\epsilon_{\mu \, \mathrm{iso}}^{(\cPqb)} = 0.029^{+0.003}_{-0.002}$ and
$\epsilon_{\Pe \, \mathrm{iso}}^{(\cPqb)} = 0.036^{+0.013}_{-0.008}$, with uncertainties due to statistics only.

We probed four different potential
sources of systematic uncertainties in the
{BTag-and-probe} method, and added their contributions in quadrature
to arrive at a total systematic uncertainty on the $\epsilon_{iso}^{(\cPqb)}$ efficiency of 54 (29)\% for electrons (muons).
The largest contribution to this uncertainty is due to
the statistical precision with which the method in simulation is verified.
The subdominant contributions include
the change in the background prediction
when the $H_T$ requirement for selecting the $\bbbar$ control
sample in data is varied from 100 to 150 GeV, or when the W+jets contribution is assumed twice the size predicted
by the simulation with a value of $\epsilon^{(\cPqb)}_{iso}$ that differs from $\ttbar$ by a factor two,
and the effect of reweighting the simulated events to
lower the fraction of leptons not from b decay in the sample used to measure
$\epsilon^{(\cPqb)}_{iso}$.
We then multiply the {sideband} yield in each of the four channels $\mu\mu, \Pe\Pe, \Pe\mu$, and $\mu\Pe$
by the appropriate factor $\epsilon_{iso}^{(\cPqb)} / ( 1 - \epsilon_{iso}^{(\cPqb)} )$ to arrive at
$0.52 \pm 0.24 \pm 0.26$
as the estimate of the contribution of events with one fake lepton to the total background.
The uncertainties quoted are statistical and systematic, respectively, taking correlations into account.

While the {BTag-and-probe} technique described above is based on different assumptions than the
TL method of Section~\ref{sec:leptrg},
we note that in their implementation the two techniques are quite similar. Both techniques use events
in a \RelIso sideband combined with a scale factor determined from an independent control sample to
estimate the background in the signal region. The most notable differences are the requirement of the away-jet b-tag,
which targets leptons from b decay in the {BTag-and-probe} method,
the choice of variables used to parametrize $\epsilon_{T/L}$ in one case
and $\epsilon^{(\cPqb)}_{\mu\ iso}$ or $\epsilon^{(\cPqb)}_{\Pe\ iso}$ in the other,
and the size of the \RelIso sideband used in the extrapolation.

Combining the background estimates for events with one and two fake leptons
and propagating all statistical and systematic uncertainties between channels,
including their correlations,
we arrive at a final estimate of background due to fake leptons of $0.70 \pm 0.23 \pm 0.21 $
events, with the first (second) uncertainty being statistical (systematic).

\subsection{Search using Hadronic Triggers and $\tau_h$}
\label{sec:taufake}

Simulation studies show clearly that the largest source of background for the $\tau_h$\ channels is due to $\tau_h$ fake leptons.
We estimate this background using the same "Tight-Loose" (TL) method as was used for fake leptons in
Section~\ref{sec:leptrg}, except that for the "Loose" selection
we loosen the $\tau_h$\ identification instead of the isolation. To be specific,
part of the discrimination between hadronic $\tau$\ decays
and generic QCD jets is based on five neural networks
trained to identify different hadronic $\tau$\ decay modes.
The neural network requirements are used for the tight, but not the loose selection.

As in Section~\ref{sec:leptrg},
in order to predict the number of events from fake $\tau_h$,
we measure the tight-to-loose ratio $\epsilon_{TL}$
in bins of  $\eta $ and $p_T$.
On average, $\epsilon_{TL}=9.5\pm 0.5$\%, where the uncertainty is statistical only.
This is measured using a single-$\tau_h$ control sample
with  $H_T$\  $>$ 300~GeV and \met\ $<$ 20 GeV.
The $H_T$ requirement results in
hadronic activity similar to our signal region,
while the \met\ requirement
reduces contributions from
W, Z plus jets, and $\ttbar$,
resulting in a control region that is dominated by QCD multijet production.

\begin{table}[!h]
\caption{ \label{tab:tauFR_vali_MC}
Validation of the TL method. The number of observed events is compared to the number of predicted events in
simulation (first two columns) and in a background-dominated control region with relaxed selection criteria (last two columns).
The simulation is normalized to 35 pb$^{-1}$.
The first and second uncertainties in the number of predicted events in data are statistical and  systematic, respectively.
}
  \begin{center}
    \begin{tabular}{|l||c|c||c|c|}
\hline
 &\multicolumn{2}{c||}{Simulation} &\multicolumn{2}{c|}{ Data }\\
 &\multicolumn{2}{c||}{Only SM} &\multicolumn{2}{c|}{Relaxed selection }\\
\hline
Channel & Observed & Predicted     &  Observed & Predicted\\
\hline
$\tau \tau$     &   0.08$\pm$0.03   &   0.15$\pm$0.15 &   14 &14.0$\pm$4.3$\pm$2.6\\
\hline
$\Pe \tau$     &  0.35$\pm$0.12   &   0.30$\pm$0.11 & 1 &0.8$\pm$0.4$\pm$0.1\\
\hline
$\mu \tau$     & 0.47$\pm$0.15  &  0.49$\pm$0.20    & 2 &2.9$\pm$0.6 $\pm$0.4\\
\hline
\end{tabular}
\end{center}
\end{table}

The expected number of background events is estimated by selecting $e\tau$, $\mu\tau$, and $\tau\tau$ events where the $\tau$ candidates
pass the loose selection but fail the tight selection. We find 1, 2, and 2 such events, respectively, in these three channels.
For the $\Pe\tau$ and $\mu\tau$ channels, these events are weighted by the corresponding factors of
$\epsilon_{TL}/(1-\epsilon_{TL})$, while for the $\tau\tau$ channel each event is weighted by the product of
two such factors, one corresponding to each of the two $\tau$ leptons in the event.

We perform two types of validation of this background estimate.
First, we compare observation and prediction in the signal region for simulation.
Second, we compare observation and prediction in data after relaxing the  $H_T$\  selection from 350 GeV to 150 GeV, and removing the \met\ requirement.
In simulation the contribution of LM0 represents less than 4\% of events with two same-sign isolated leptons and $H_T$\ $>$ 150 GeV.
Table~\ref{tab:tauFR_vali_MC} presents both of these validations of the background estimation technique.
We find good agreement between the observation and prediction in all channels in data and in the simulation.

The largest source of systematic uncertainties
in the prediction of background events is due to lack of statistics of simulated events to validate the method (30\%).
In addition, we find uncertainties of
18\%, 8\% and 7\% in $\tau_h\tau_h$, $\Pe\tau_h$, and $\mu\tau_h$, respectively, due to
the correlation of $\epsilon_{TL}$ with $H_T$.
We measure $\epsilon_{TL}$ for $H_T > 150$\ GeV. We determine the systematic
uncertainty as the difference in the
number of predicted background events from the reference measurement at $H_T > 300$ GeV with the measurement for $H_T > 150$ GeV.
From simulation studies it is found that an
additional 10\% systematic uncertainty must be added in quadrature in the $\Pe\tau_h$ and $\mu\tau_h$ channels
to account for neglecting background contributions from fake electrons and muons.
Taking all of this into account, we arrive at an estimate of $0.28 \pm 0.14 \pm 0.09$ events with fake leptons,
where the uncertainties are statistical and systematic, respectively.

\subsection{Comparison of Leptonic and $H_T$-Triggered Analyses with Electrons and Muons in the Final State}
\label{sec:compare}

As discussed above, we utilize two different trigger strategies to define selections that cover the maximum phase space
possible in our search for new physics. 
In addition, the fact that these two selections have an overlap allows us to perform direct comparisons and cross-checks
that we present in this section.

We start by defining an overlap preselection requiring one electron or muon of $p_T > $20 GeV and a second with $p_T > $ 10 GeV,
$H_T > $ 300 GeV, and no \met or isolation requirements. We note that this corresponds to the {preselection} sample
from Section~\ref{sec:Florida} with the $p_T$ of the leptons tightened to be consistent with Section~\ref{sec:leptrg}.
Comparing yields from this selection for the two trigger strategies 
on an event-by-event basis, we find that all of the $H_T$-triggered events are also
present in the lepton-triggered sample. 
From this comparison, we calculate efficiencies for the $H_T$ and leptonic triggers to be
 $(92\pm 4)$ \% and $(100^{+0}_{-2})$ \%, respectively.
While statistics are limited, this confirms the trigger efficiencies
measured in independent data samples, as presented in Section~\ref{sec:sr}.

Sections~\ref{sec:leptrg} and \ref{sec:Florida} introduced two
  alternative methods for estimating the background due to fake
  leptons.  Here we compare the two independent predictions
  in the region of overlap for the two searches.
To compare the two methods in a common signal region, we require  
\met $> 30$ GeV and the lepton isolation described in Section~\ref{sec:leptrg}, in addition to the preselection defined above.
The TL method predicts $0.68 \pm 0.39$ based on a yield of 3 events that pass the loose selection.
The second method introduced in Section~\ref{sec:Florida} results in an estimate of $0.27 \pm 0.12$ events based on 11 events in the
\RelIso sideband. Both of these uncertainties are statistical only.
We thus conclude that the two trigger strategies lead to consistent results within the kinematic region where they overlap, and
the two methods of estimating backgrounds due to fake leptons give consistent results in that region.

\subsection{Electron Charge Mismeasurement}
\label{sec:flip}

A second potentially important source of background consists
of opposite-sign dilepton events ($\Pe^{\pm}\Pe^{\mp}$ or $\Pe^{\pm}\mu^{\mp}$)
where the sign of the charge of one of the electrons is mismeasured
because of hard bremsstrahlung in the tracker volume.

We measure the electron charge in three different ways.
Two of the measurements are based on
the reconstructed track from two separate tracking algorithms: the standard
CMS track reconstruction algorithm~\cite{TRK-10-001,Adam:2005cg} and the Gaussian Sum Filter algorithm~\cite{ref:GSF},
optimized
for the measurement of electron tracks that radiate in the tracker
material.  The third measurement is based on the relative position of the
calorimeter cluster and the projection to the calorimeter of
a line segment built out of hits in the pixel detector.
To reduce the effect of charge mismeasurements, we require
agreement among the three measurements.

\begin{figure}[tbh]
\begin{center}
\includegraphics[width=1.0\linewidth]{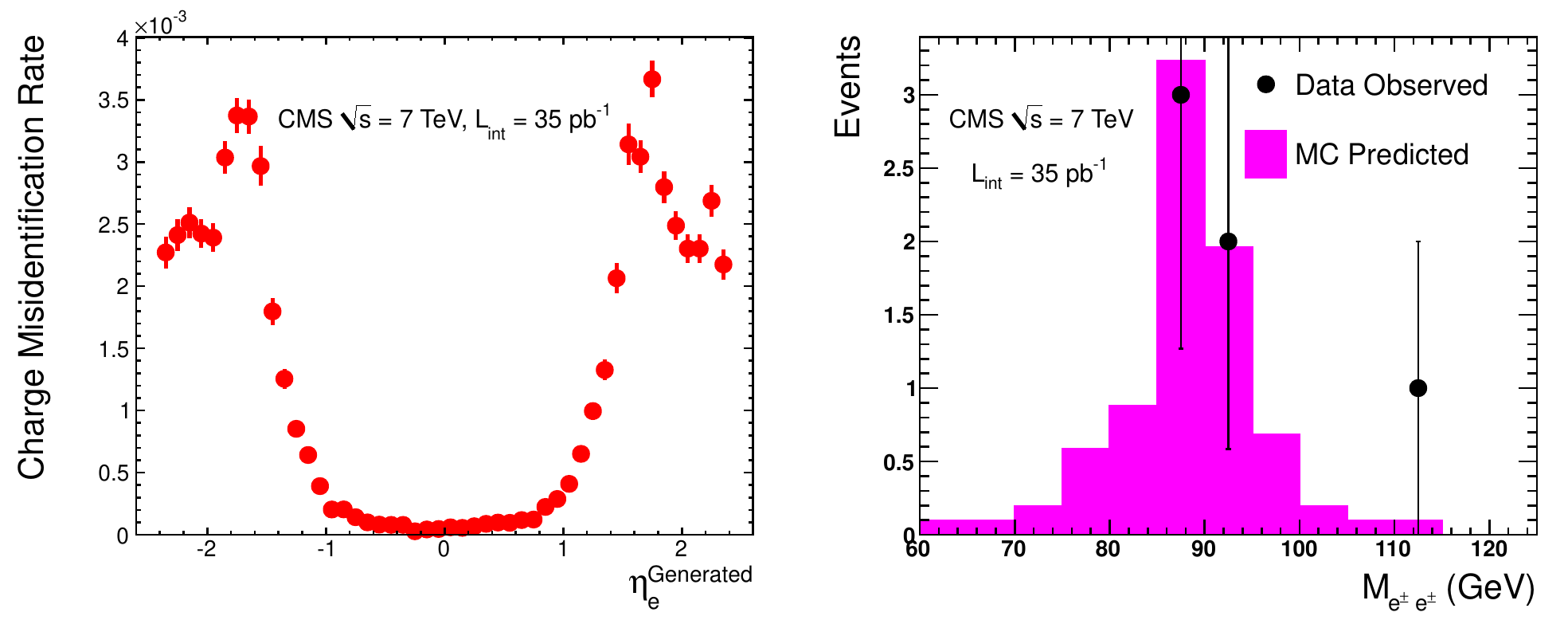}
\caption{\label{fig:fliprate}\protect (Left) The probability to mismeasure
the electron charge as a function of $\eta$ in the $p_T$ range
10$-$100 GeV, as obtained from simulation.
(Right) Same-sign $\Pe\Pe$ invariant mass
distribution in data compared with the $\cPZ \to \Pe\Pe$ expectation from simulation.
}
\end{center}
\end{figure}

After this requirement, the probability of mismeasuring the charge
of an electron in simulation is at the level of a few per mille, even in the $|\eta|>1$
region where the amount of material is largest, as can be seen in Fig.~\ref{fig:fliprate} (left).

To demonstrate our understanding
of this probability, we show in Fig.~\ref{fig:fliprate} (right) the invariant-mass
spectrum for same-sign $\Pe\Pe$ events and our simulation-based prediction
from $\cPZ \to \Pe\Pe$ with one mismeasured charge.
The $\cPZ$ sample shown uses the tight electron selection described in Section~\ref{sec:srlep},
except with no jet or $H_T$ requirement.
Instead, we require
\met\ $<$ 20 GeV and transverse mass
$<$ 25 GeV to reduce backgrounds from $\PW +$ jets.
The highest-$p_T$ lepton has been used in the calculation of the transverse
mass.

Measurement of the electron momentum is dominated by the energy measurements in the calorimeter, while the measurement
of its charge is dominated by measurements in the tracker. An electron with mismeasured charge will thus still
have a correctly measured momentum,
leading to the clear $\cPZ$ peak in the $\Pe^\pm \Pe^\pm$ invariant-mass displayed in Fig.~\ref{fig:fliprate} (right).
Normalized to 35 pb$^{-1}$, the simulation predicts $7.4\pm 0.9$ events in the $\cPZ$ mass region, with
the quoted uncertainty due to statistics. In data, we observe 5 events in the same region.
We predict the same-sign $\cPZ$ yield in
simulation ($6.37\pm 0.03$) and data ($4.9 \pm 0.1$)
based on reweighting the opposite-sign $\cPZ\to \Pe^\pm \Pe^\mp$ yield by a simulation-based parametrization of the
probability for electron charge mismeasurement as a function of $p_T$ and $\eta$.

For the leptonic trigger searches we estimate the number of background events due to charge misidentification by scaling the
opposite-sign yields by the above probability function.
We estimate the background due to electron charge misidentification as
$0.012\pm 0.002$ and
$0.04\pm 0.01$ for the 
\met\ $>$ 80 GeV and $H_T > $ 200 GeV regions, respectively.

For the $H_T$-triggered searches these backgrounds are further reduced since the opposite-sign yield is smaller given the
tighter $H_T$ requirement.
The resulting background prediction is $0.008 \pm 0.005$ ($\Pe\Pe$) and $0.004 \pm 0.002$ ($\Pe\mu $) events.
For the search with $\tau_h$ in the final state, this background is negligible and ignored, as even in the
opposite-sign $\Pe\tau_h$
channel, background $\tau_h$ contributions dominate over those with a prompt $\tau_h$.

We assign a 50\% systematic uncertainty on the estimated backgrounds due to electron charge mismeasurement.
This is motivated by the statistics available in the doubly charged $\cPZ \to \Pe^\pm \Pe^\pm$ signal region.

\section{Signal Acceptance and Efficiency Systematic Uncertainties}
\label{sec:systematics}

Electron and muon identification efficiencies above $p_T \approx 20$ GeV are known at the
level of 3\% per electron and 1.5\% per muon, based on studies of large samples of $\cPZ\to\Pe\Pe$ and $\mu\mu$ events in data and simulation.
The uncertainties increase as the
efficiencies themselves decrease towards lower $p_T$, reaching 6\% (8\%) per muon (electron) at 5 (10) GeV,
based on studies of large samples of
$\cPZ \to \Pe\Pe$ and $\mu\mu$ in data and simulation.
In addition, there is a potential mismodelling of the lepton isolation efficiency between data and simulation
that grows with the amount of hadronic activity per event.
To assess this, we compare the isolation efficiency as a function of track multiplicity in data and simulation for
$\cPZ \to \Pe\Pe$ and $\mu\mu$, and extrapolate to new physics signals with large hadronic activity using
simulation, as discussed in more detail in Section~\ref{sec:discussion}. Based on this, we assign an additional
5\% systematic uncertainty per lepton.
There is also a
1\% (5\%) uncertainty associated with the lepton ($H_T$) trigger efficiency.

The efficiency of the hadronic $\tau_h$ selection is studied in data
via the process
$\cPZ\rightarrow\tau \tau$, where one $\tau$ decays hadronically while the other decays into a muon~\cite{PFT-10-004}.
The available statistics are an order of magnitude lower than the statistics available in $\cPZ\to\Pe\Pe$ or $\mu\mu$,
at significantly lower purity.
Accordingly, $\tau_h$\ reconstruction versus $p_T$
can not be studied at the same level of detail in data as for the electron and muon reconstruction,
and we depend to a greater extent on an accurate simulation than we do for electrons and muons.
We assign an uncertainty of 30\%~\cite{PFT-10-XXX} to the $\tau_h$ selection efficiency
to account for the limited statistics available in data to validate the efficiency measured in simulation.

An additional source of systematic uncertainty is associated with the current
$\approx 5\%$ uncertainty on the hadronic energy scale~\cite{JES} at CMS.
This scale uncertainty limits our understanding of the
efficiency of the $H_T$ and \met requirements.
Clearly, final states where the typical $H_T$ and \met are large
compared to the selection values used in the analysis are less
affected than those with smaller $H_T$ and \met.
We compute the systematic uncertainty due to this effect
for the LM0 benchmark point with the four signal selections
using the method of Ref.~\cite{ref:top}.
We use the LM0 model as it is typical
of the possible SUSY final states to which these analyses
are sensitive.  We find that the uncertainty varies
between 1\% at $H_T > $ 60 GeV
and 7\% at $H_T > $ 350 GeV, the values of $H_T$ used
in the selections for the lepton-triggered baseline and $\tau_h$ search respectively.

Uncertainties in the acceptance due to the modelling of initial- and final-state radiation and knowledge of
the parton density functions (PDF) are estimated to be 2\%. For the latter, we use the CTEQ6.6~\cite{cteq66} PDF and
their uncertainties.

Based on LM0 as a signal model, we arrive at total uncertainties on signal efficiencies of 12\%, 15\%, and 30\% for the lepton triggered,
$H_T$ triggered low $p_T$, and $H_T$ triggered $\tau_h$ analyses, respectively.
This includes a 4\% luminosity systematic uncertainty~\cite{lumi}.
In addition, to interpret these limits in terms of constraints on new physics models, one needs to take into account
any model-dependent theoretical uncertainties.

\section{Summary of Results}
\label{sec:results}

\begin{table}[htb]
\begin{center}
\caption{\label{tab:summary}
Observed and estimated background yields for all analyses.
The rows labeled ``{\bf predicted BG}'' refer to the sum of the data-driven estimates of the fake lepton contributions,
and the residual contributions predicted by the simulation. The rows labeled ``MC'' refer to the background as predicted from
the simulation alone.
Rows labeled ``{\bf observed}'' show the actual number of events seen in data.
The last column (95\% CL UL Yield) represents observed upper limits on event
yields from new physics.}
\begin{tabular}{|c|c|c|c|c|c|}
\hline
Search Region & $\Pe\Pe$    & $\mu\mu$ & $\Pe\mu$      & total  & 95\% CL UL Yield \\ \hline
\hline
Lepton Trigger & & & & & \\
\hline
\met\ $> 80$\ GeV & & & & &\\
MC & 0.05 & 0.07 & 0.23 & 0.35 & \\
{\bf predicted BG} & {\boldmath $0.23^{+0.35}_{-0.23}$} & {\boldmath $0.23^{+0.26}_{-0.23}$} & {\boldmath $0.74\pm 0.55$} & {\boldmath $1.2\pm 0.8$} & \\
{\bf observed}  & {\bf 0} & {\bf 0} & {\bf 0} & {\bf 0}  & {\bf 3.1} \\
\hline
$H_T > 200$\  GeV & & & & &\\
MC & 0.04 & 0.10 & 0.17 & 0.32 &  \\
{\bf predicted BG} & {\boldmath $0.71\pm 0.58$} & {\boldmath $0.01^{+0.24}_{-0.01}$} & {\boldmath $0.25^{+0.27}_{-0.25}$} &
{\boldmath $0.97\pm 0.74$} &  \\
{\bf observed}  & {\bf 0} & {\bf 0} & {\bf 1} & {\bf 1}  & {\bf 4.3} \\
\hline
\hline
$H_T$\ Trigger & & & & & \\
\hline
Low-$p_T$
& & & & &\\
 MC & 0.05 & 0.16 & 0.21 & 0.41 &  \\
 {\bf predicted BG} &  \boldmath{ $0.10\pm 0.07$} & {\boldmath  $0.30\pm 0.13$ } & \boldmath{$0.40\pm 0.18$} & \boldmath{$0.80\pm 0.31$} &  \\
 {\bf observed } &{\bf  1} & {\bf 0} & {\bf 0 } & {\bf 1} & {\bf 4.4} \\
\hline
& $\Pe\tau_h$    & $\mu\tau_h$ & $\tau_h\tau_h$      & total  & 95\% CL UL Yield \\ \hline
$\tau_h$\ enriched& & & & &\\
 MC & 0.36 & 0.47 & 0.08 & 0.91 & \\
{\bf predicted BG} &{\boldmath  $0.10\pm 0.10$ }& \boldmath{ $0.17\pm 0.14$} & \boldmath{$0.02\pm 0.01$} &  \boldmath{$0.29\pm 0.17$} & \\ 
 {\bf observed } &{\bf  0} & {\bf 0} & {\bf 0 } & {\bf 0} & {\bf 3.4} \\
\hline
\end{tabular}
\end{center}
\end{table}

\begin{figure}[htbp]
\begin{center}
\includegraphics[angle=0,width=0.60\textwidth]{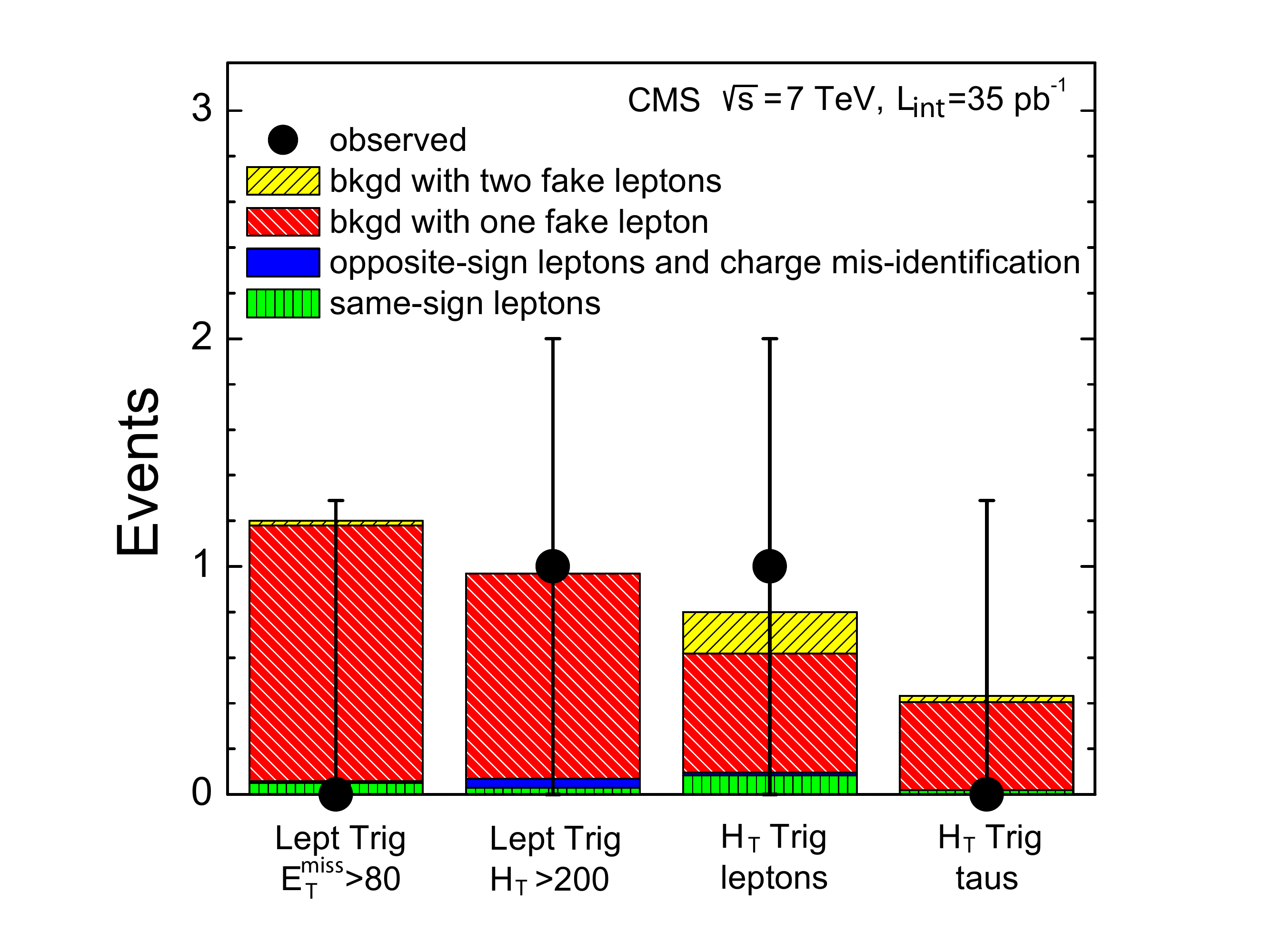}
\caption{
\label{fig:barchart}
A visual summary of the observed number of data events, the expected number of background events, and the
composition of the background for the four search regions.
}
\end{center}
\end{figure}

The results of our searches are summarized in Table~\ref{tab:summary}.
The background (BG) predictions are given by the rows labelled "predicted BG".
In addition to the background estimate from data, we also present an
 estimate of the background based on simulation in the rows labeled as "MC".
While QCD multijet production samples are used for testing background estimation methods
in our control regions, they are too statistically limited to provide meaningful estimates of
yields in the signal regions listed in Table~\ref{tab:summary}, and are thus not included. All other
SM simulation samples described in Section~\ref{sec:sr} are represented.
Figure~\ref{fig:barchart} summarizes the signal region yields and background composition
in all four search regions presented in Table~\ref{tab:summary}.
The lepton plus jets background
where the second lepton candidate is a fake lepton from a
jet clearly dominates all search regions.
The low-$p_T$-lepton analysis has a small, but non-negligible, background contribution from
events with two fake leptons.
Estimates for backgrounds due to events with one or two fake leptons
were obtained directly from data
in appropriately chosen control regions, as described in detail in
Sections~\ref{sec:leptrg}, \ref{sec:Florida}, and \ref{sec:taufake}.
In the $\Pe\Pe$ and $\Pe\mu$ final states, small additional background constributions are present
due to the electron charge mismeasurement, as discussed in Section~\ref{sec:flip}.
The remaining
irreducible background from two prompt isolated
same-sign leptons (WZ, ZZ, $\ttbar\PW$, etc.)
amounts to at most 10\% of the total
and is estimated based on theoretical cross section predictions and simulation.
Uncertainties on the background prediction include statistical and systematic uncertainties added in quadrature.
Contributions estimated with simulation are assigned a 50\% systematic uncertainty. Data-driven estimates
are assigned a systematic uncertainty between 30\% and 50\% across the various signal regions and channels.
The $\Pe\Pe,\Pe\mu $, and $\mu\mu$ channels have partially or fully correlated systematic uncertainties, as
described in detail in Section~\ref{sec:bkg}.

We see no evidence of an event yield in excess of the background prediction
and set 95\% CL upper limits (UL) on the number of observed events using
a Bayesian method~\cite{pdg} with a flat prior on the signal strength and log-normal priors for efficiency and background uncertainties.
These include uncertainties on the signal efficiency of 12\% , 15\%, and 30\% for the lepton triggered,
$H_T$ triggered low-$p_T$, and $H_T$ triggered $\tau_h$ analyses, respectively, as discussed in more detail
in Section~\ref{sec:systematics}.
Based on the LM0 benchmark model, the simulation predicts 7.3, 9.6, 9.1, and 2.0 events
for the four signal regions, respectively.
These LM0 yields are based on the individual NLO cross sections for all production processes that contribute
to the expected signal yield.

\section{Interpretation of Results}
\label{sec:discussion}

One of the challenges of signature-based searches
is to convey information in a form that can be used to
test a variety of specific physics models.
In this section we present additional information
that can be used to confront models of new physics
in an approximate way by
generator-level simulation studies that
compare the expected number of events in 35 pb$^{-1}$
with our upper limits shown in Table~\ref{tab:summary}.

The kinematic requirements described in
Section~\ref{sec:sr}
are the first key ingredients of such studies.
The $H_T$ variable can be approximated by defining it
as the scalar sum of the $p_T$ of all final-state quarks (u, d, c, s,
and b) and gluons with $p_T >$ 30 GeV produced in the hard-scattering process.  The \met can be defined
as the magnitude of the vector sum of the transverse momentum over all non-interacting particles,
{\em e.g.}, neutrinos and LSP.
The ratio of the mean detector responses for $H_T$ and \met as defined above, to their true values
are $0.94 \pm 0.05$, and
$0.95 \pm 0.05$, respectively, where the uncertainties are dominated by
the jet energy scale uncertainty.  The resolution on these two
quantities differs for the different selections. In addition, the \met\ resolution
depends on the total hadronic activity in the event.
It ranges from about 7 to 25 GeV for events with $H_T$ in the range of 60 to 350 GeV.
The $H_T$ resolution decreases from about 26\% at 200 GeV to 19\%  for 300 GeV and to 18\% for 350 GeV.
The $H_T$ resolution was measured in simulation using the LM0 reference model, while
the \met resolution was measured in data. 

\begin{figure}[htbp]
\begin{center}
\includegraphics[angle=0,width=0.45\textwidth]{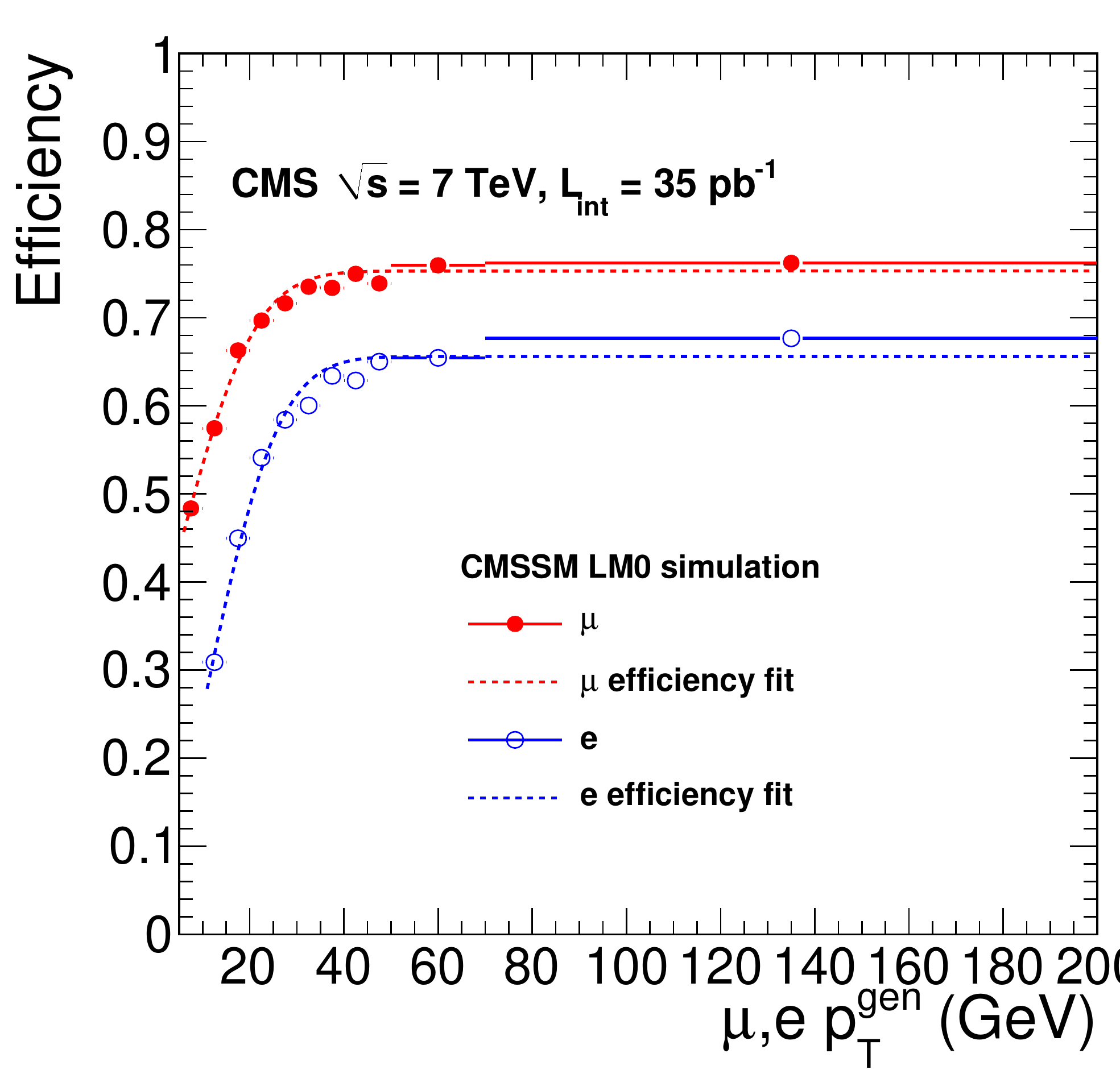}
\includegraphics[angle=0,width=0.45\textwidth]{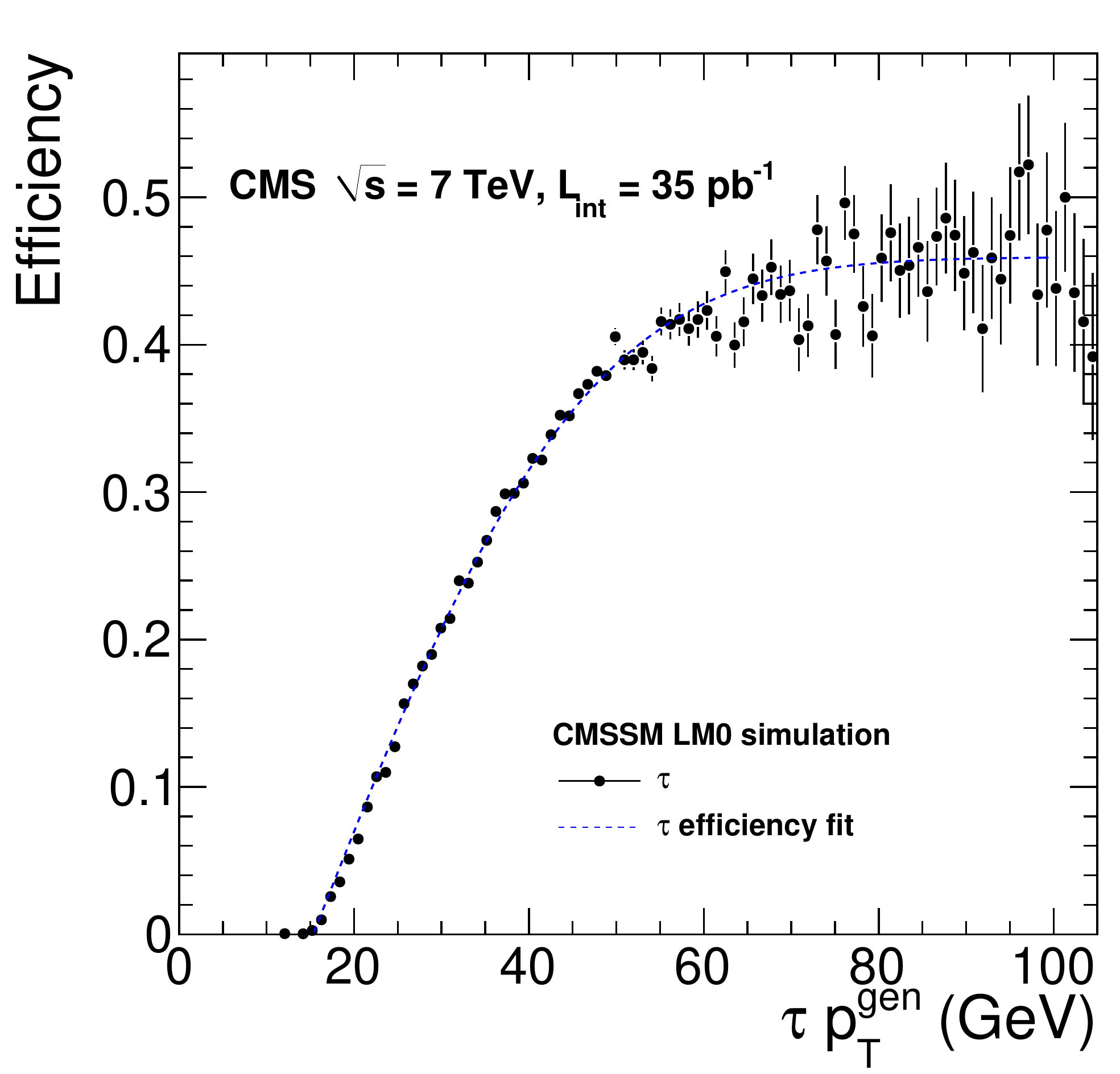}
\caption{
\label{fig:efficiency}
Electron, muon (left) and $\tau_h$ (right) selection efficiencies as a function of $p_T$.
The results of the fits described in the text are shown by the dotted lines.
}
\end{center}
\end{figure}

Figure~\ref{fig:efficiency} shows the efficiency versus $p_T$ using the LM0 reference model for $\Pe,\mu $ (left),
and $\tau_h$ (right). Efficiencies here include reconstruction, isolation, and selection.
We fit the curves in Fig.~\ref{fig:efficiency} to
the functional form: efficiency($p_T$) = $\epsilon_{max} + A\times $\ (erf(($p_T - P_{Tcut}$)$/B$) $-$1 ).
We fix $P_{Tcut}$ to 10, 5, 15 GeV, and
find ($A$, $B$, $\epsilon_{max}$) of
(0.40, 18, 0.66),
(0.32, 18, 0.75),
and (0.45, 31, 0.45) for $\Pe$, $\mu$, and $\tau_h$, respectively.

Lepton isolation efficiencies depend on the
hadronic activity in the event, and in some extreme cases like significantly  boosted top quarks, on the event topology.
The number of charged particles at the generator level after fragmentation and hadronization is a good measure
of the first of these two effects, and has been shown to agree
reasonably well between data and simulation~\cite{QCD-10-004pub}.
We find that the isolation efficiency reduces roughly linearly by 10\% for every 15 charged particles with $p_T >$ 3 GeV
within the detector acceptance. This was studied using Drell-Yan and LM0 simulation, and compared with $\mathrm{Z}\rightarrow \ell^+\ell^-$ data.
The linearity was thus shown to be valid within a range
of charged multiplicities from about 10 to 40.
The LM0 reference model shown in Fig.~\ref{fig:efficiency}
has an average charged multiplicity of $\sim$ 25 for events that pass our selections.
To arrive at the lepton efficiency for a new physics model with an average charged multiplicity of $\sim 40$,
one would take the efficiency parametrization depicted in Fig.~\ref{fig:efficiency} and multiply it by 0.9
to account for the 10\% change in isolation efficiency due to the larger average charged multiplicity.
The second effect, i.e., topologies as in boosted top quarks, is difficult to model at generator level,
as the isolation efficiency may vary by an order of magnitude or more in extreme cases.
Our results are thus not easily interpretable in new physics models with such characteristics without a detailed detector simulation.

\begin{figure}[htbp]
\begin{center}
\includegraphics[angle=0,width=0.9\textwidth]{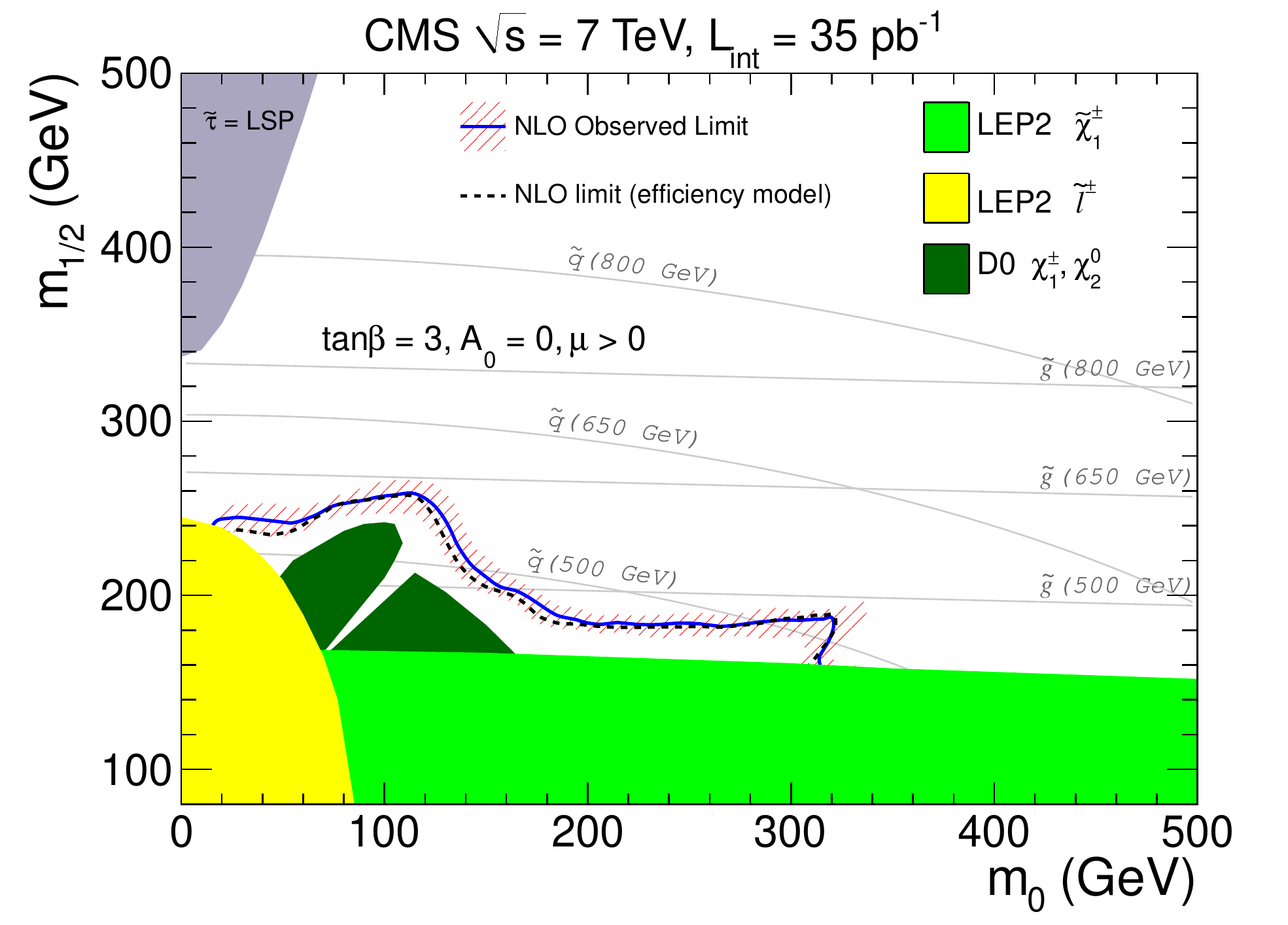}
\caption{
\label{fig:scan}
Exclusion contour in the $m_0$---$m_{1/2}$ plane for CMSSM as described in the text.
Comparing the width of the red shaded band (theoretical uncertainty) around the blue curve with the difference between
the solid blue and dashed black curves shows that the imperfections in the simple efficiency model described in the text
are small compared to the theoretical uncertainties.
}
\end{center}
\end{figure}

To show the level of precision obtainable based on such a simple efficiency model, Fig.~\ref{fig:scan}
compares the
exclusion line for CMSSM with tan$\beta = 3$, $A_0 = 0$ GeV, and $\mu >$ 0, as obtained from simulation (solid blue)
with the
corresponding curve (dashed black) for this simple efficiency model. For the exclusion plane, the SUSY particle
spectrum is calculated using SoftSUSY~\cite{softsusy} along with sparticle decay using SDECAY~\cite{sdecay}. The signal events are generated
with PYTHIA 6.4.22~\cite{pythia} using CTEQ6m~\cite{cteq} PDF. The NLO cross sections are obtained with the cross section calculator
Prospino~\cite{prospino} at each point in the $(m_0,m_{1/2})$ plane individually.

When applying the simple efficiency model,
we use the description above for the \met\ and $H_T$ response and resolution, and the efficiency functions displayed
in Fig.~\ref{fig:efficiency}. No attempt was made here to correct for differences in hadronic event environment
as those are small for the relevant $(m_0,m_{1/2})$ points.
The width of the red shaded band around the blue line indicates
uncertainties in the NLO cross section calculation. It includes variations of the PDF and
simultaneous variation by a factor of two of the renormalization and factorization scales.
Both effects are added in quadrature.
Figure~\ref{fig:scan} shows that the theoretical uncertainties are larger than the imperfections in the simple efficiency model.
The specific limit shown here corresponds to the leptonic trigger
result with \met $ > $ 80 GeV for the purpose of illustration.
The contour separates the bottom-left region where
the expected event yield would be larger than the observed limit of 3.1 events (see Table~\ref{tab:summary}, first row, last column)
 from the top-right region where such expected yield would be lower.

We choose this CMSSM model because it provides a common reference point to compare with
previously published Tevatron results~\cite{CDFtrilep, Abazov200934}.
The excluded regions from LEP are based on searches for sleptons and charginos~\cite{leplimit,aleph,delphi,l3,opal}.
Other final states, especially the all-hadronic~\cite{SUS-10-003}, are better suited for this
standard reference model, while the leptonic same-sign final state explored in this paper is more appropriate
to constrain a wide variety of other
new physics models~\cite{Barnett:1993ea,Guchait:1994zk,Baer:1995va,Cheng:2002ab,Contino:2008hi,Almeida:1997em,Han:2009}.
As discussed in Section~\ref{sec:sr},
in a general supersymmetry context, one might expect gluino-gluino or gluino-squark production to lead to same-sign
dilepton events via a decay chain involving a chargino. The salient, and very generic feature here is one lepton per gluino with
either sign being equally likely. As a result, 50\% of the dilepton events will be same-sign.
Moreover, these cascade decays are typically characterized by two mass difference scales that separately determine typical
$H_T$ and lepton $p_T$ values. Different models of new physics may thus populate only one or the other of our different
search regions.

\section{Summary and Conclusions}

Using two different trigger strategies,
we have searched for new physics with same-sign dilepton events in the $\Pe\Pe$,
$\mu\mu$, $\Pe\mu$, $\Pe\tau_h$, $\mu\tau_h$, and $\tau_h\tau_h$ final states,
and have seen no evidence for an excess over the background prediction. The $\tau_h$ leptons referred to here are
reconstructed via their
hadronic decays.
The dominant background processes in all final states except $\tau_h\tau_h$ involve events with one fake lepton.
In the $\tau_h\tau_h$ final state, events with two fake $\tau_h$ dominate.
We have presented methods to derive background estimates from the data for all major background sources.
We have set 95\% CL upper limits
on the number of signal events within $|\eta | < 2.4$
at 35 pb$^{-1}$ in the range of 3.1 to 4.5 events, depending on signal region,
and have presented details on signal efficiencies
that can be used to confront a wide variety of models of new physics.
Our analysis extends the region excluded by experiments at LEP and the Tevatron in the CMSSM model.

\section{Acknowledgements}
We wish to congratulate our colleagues in the CERN accelerator departments for the excellent performance of the LHC machine. We thank the technical and administrative staff at CERN and other CMS institutes, and acknowledge support from: FMSR (Austria); FNRS and FWO (Belgium); CNPq, CAPES, FAPERJ, and FAPESP (Brazil); MES (Bulgaria); CERN; CAS, MoST, and NSFC (China); COLCIENCIAS (Colombia); MSES (Croatia); RPF (Cyprus); Academy of Sciences and NICPB (Estonia); Academy of Finland, MEC, and HIP (Finland); CEA and CNRS/IN2P3 (France); BMBF, DFG, and HGF (Germany); GSRT (Greece); OTKA and NKTH (Hungary); DAE and DST (India); IPM (Iran); SFI (Ireland); INFN (Italy); NRF and WCU (Korea); LAS (Lithuania); CINVESTAV, CONACYT, SEP, and UASLP-FAI (Mexico); MSI (New Zealand); PAEC (Pakistan); SCSR (Poland); FCT (Portugal); JINR (Armenia, Belarus, Georgia, Ukraine, Uzbekistan); MST and MAE (Russia); MSTD (Serbia); MICINN and CPAN (Spain); Swiss Funding Agencies (Switzerland); NSC (Taipei); TUBITAK and TAEK (Turkey); STFC (United Kingdom); DOE and NSF (USA).

Individuals have received support from the Marie-Curie programme and the European Research Council (European Union); the Leventis Foundation; the A. P. Sloan Foundation; the Alexander von Humboldt Foundation; the Associazione per lo Sviluppo Scientifico e Tecnologico del Piemonte (Italy); the Belgian Federal Science Policy Office; the Fonds pour la Formation \`a la Recherche dans l'Industrie et dans l'Agriculture (FRIA-Belgium); the Agentschap voor Innovatie door Wetenschap en Technologie (IWT-Belgium); and the Council of Science and Industrial Research, India.

\clearpage
\bibliography{auto_generated}   

\providecommand{\href}[2]{#2}\begingroup\raggedright\begin{thebibliography}{10}%
\makeatletter
\providecommand{\hrefCMSnoop }[0]{\@secondoftwo}%
\makeatother

\bibitem{Barnett:1993ea}
\hrefCMSnoop {} {R.~Barnett, J.~Gunion, and H.~Haber, ``Discovering
  supersymmetry with like sign dileptons'',} \textit{ Phys.\ Lett.} \textbf{ B
  315} (1993) 349.
  \href{http://dx.doi.org/10.1016/0370-2693(93)91623-U}{\texttt{
  doi:10.1016/0370-2693(93)91623-U}}.

\bibitem{Guchait:1994zk}
\hrefCMSnoop {} {M.~Guchait and D.~P. Roy, ``Like sign dilepton signature for
  gluino production at the CERN LHC including top quark and Higgs boson
  effects'',} \textit{ Phys.\ Rev.} \textbf{ D 52} (1995) 133.
  \href{http://dx.doi.org/10.1103/PhysRevD.52.133}{\texttt{
  doi:10.1103/PhysRevD.52.133}}.

\bibitem{Baer:1995va}
\hrefCMSnoop {} {H.~Baer {et~al.}, ``Signals for minimal supergravity at the
  CERN Large Hadron Collider II: Multilepton channels'',} \textit{ Phys.\ Rev.}
  \textbf{ D 53} (1996) 6241.
  \href{http://dx.doi.org/10.1103/PhysRevD.53.6241}{\texttt{
  doi:10.1103/PhysRevD.53.6241}}.

\bibitem{Cheng:2002ab}
\hrefCMSnoop {} {H.~Cheng, K.~Matchev, and M.~Schmaltz, ``Bosonic
  supersymmetry? Getting fooled at the CERN LHC'',} \textit{ Phys.\ Rev.}
  \textbf{ D 66} (2002) 056006.
  \href{http://dx.doi.org/10.1103/PhysRevD.66.056006}{\texttt{
  doi:10.1103/PhysRevD.66.056006}}.

\bibitem{Contino:2008hi}
\hrefCMSnoop {} {R.~Contino and G.~Servant, ``Discovering the top partners at
  the LHC using same-sign dilepton final states'',} \textit{ JHEP} \textbf{ 06}
  (2008) 026. \href{http://dx.doi.org/10.1088/1126-6708/2008/06/026}{\texttt{
  doi:10.1088/1126-6708/2008/06/026}}.

\bibitem{Almeida:1997em}
\hrefCMSnoop {} {F.~Almeida {et~al.}, ``Same-sign dileptons as a signature for
  heavy Majorana neutrinos in hadron-hadron collisions'',} \textit{ Phys.
  Lett.} \textbf{ B 400} (1997) 331.
  \href{http://dx.doi.org/10.1016/S0370-2693(97)00143-3}{\texttt{
  doi:10.1016/S0370-2693(97)00143-3}}.

\bibitem{Han:2009}
\hrefCMSnoop {} {Y.~Bai and Z.~Han, ``Top-antitop and Top-top Resonances in the
  Dilepton Channel at the CERN LHC'',} \textit{ JHEP} \textbf{ 04} (2009) 056.
  \href{http://dx.doi.org/10.1088/1126-6708/2009/04/056}{\texttt{
  doi:10.1088/1126-6708/2009/04/056}}.

\bibitem{dm1}
\hrefCMSnoop {} {G.~Bertone, D.~Hooper, and J.~Silk, ``Particle dark matter:
  Evidence, candidates and constraints'',} \textit{ Phys. Rept.} \textbf{ 405}
  (2005) 279. \href{http://dx.doi.org/10.1016/j.physrep.2004.08.031}{\texttt{
  doi:10.1016/j.physrep.2004.08.031}}.

\bibitem{CMS}
\hrefCMSnoop {} {{ CMS} Collaboration, ``The CMS experiment at the CERN LHC'',}
  \textit{ JINST} \textbf{ 3} (2008) S08004.
  \href{http://dx.doi.org/10.1088/1748-0221/3/08/S08004}{\texttt{
  doi:10.1088/1748-0221/3/08/S08004}}.

\bibitem{MUOPAS}
\href {http://cdsweb.cern.ch/record/1279140} {{ CMS} Collaboration,
  ``Performance of muon identification in pp collisions at $\sqrt{s}$ = 7
  {TeV}'',} \textit{ CMS Physics Analysis Summary} \textbf{
  \href{http://cdsweb.cern.ch/record/1279140}{CMS-PAS-MUO-10-002}} (2010).

\bibitem{EGMPAS}
\href {http://cdsweb.cern.ch/record/1299116} {{ CMS} Collaboration, ``Electron
  reconstruction and identification at $\sqrt{s}$=7 {TeV}'',} \textit{ CMS
  Physics Analysis Summary} \textbf{
  \href{http://cdsweb.cern.ch/record/1299116}{CMS-PAS-EGM-10-004}} (2010).

\bibitem{inclusWXsect}
\hrefCMSnoop {} {{ CMS} Collaboration, ``Measurements of Inclusive W and Z
  Cross Sections in pp Collisions at $\sqrt{s}=7$ {TeV}'',} \textit{ JHEP}
  \textbf{ 01} (2011) 080.
  \href{http://dx.doi.org/10.1007/JHEP01(2011)080}{\texttt{
  doi:10.1007/JHEP01(2011)080}}.

\bibitem{PFT-10-004}
\href {http://cdsweb.cern.ch/record/1279358} {{ CMS} Collaboration, ``Study of
  tau reconstruction algorithms using $\Pp\Pp$ collisions data collected at
  $\sqrt{s} = 7\,${TeV} with {CMS} detector at {LHC}'',} \textit{ CMS Physics
  Analysis Summary} \textbf{ CMS-PAS-PFT-10-004} (2010).

\bibitem{PFT-09-001}
\href {http://cdsweb.cern.ch/record/1194487} {{ CMS} Collaboration,
  ``Particle--Flow Event Reconstruction in {CMS} and Performance for Jets,
  Taus, and {\MET}'',} \textit{ CMS Physics Analysis Summary} \textbf{
  CMS-PAS-PFT-09-001} (2009).

\bibitem{PFT-10-XXX}
\href {http://cdsweb.cern.ch/record/1337004} {{ CMS} Collaboration,
  ``Performance of tau reconstruction algorithms in 2010 data collected with
  {CMS}'',} \textit{ CMS Physics Analysis Summary} \textbf{ CMS-PAS-TAU-11-001}
  (2011).

\bibitem{PFT-10-002}
\href {http://cdsweb.cern.ch/record/1279341} {{ CMS} Collaboration,
  ``Commissioning of the Particle-Flow Reconstruction in Minimum-Bias and Jet
  Events from \Pp\Pp\ Collisions at 7 {TeV}'',} \textit{ CMS Physics Analysis
  Summary} \textbf{ CMS-PAS-PFT-10-002} (2010).

\bibitem{anti-kt}
\hrefCMSnoop {} {M.~Cacciari, G.~Salam, and G.~Soyez, ``The anti-$k_T$ jet
  clustering algorithm'',} \textit{ JHEP} \textbf{ 04} (2008) 063.
  \href{http://dx.doi.org/10.1088/1126-6708/2008/04/063}{\texttt{
  doi:10.1088/1126-6708/2008/04/063}}.

\bibitem{JME-10-001}
\href {http://cdsweb.cern.ch/record/1248210} {{ CMS} Collaboration, ``Jets in
  0.9 and 2.36 {TeV} pp Collisions'',} \textit{ CMS Physics Analysis Summary}
  \textbf{ CMS-PAS-JME-10-001} (2010).

\bibitem{JES}
\href {http://cdsweb.cern.ch/record/1279362} {{ CMS} Collaboration, ``Jet
  Performance in pp Collisions at $\sqrt{s}$=7 {TeV}'',} \textit{ CMS Physics
  Analysis Summary} \textbf{ CMS-PAS-JME-10-003} (2010).

\bibitem{cmssm}
\hrefCMSnoop {} {G.~Kane {et~al.}, ``Study of constrained minimal
  supersymmetry'',} \textit{ Phys.\ Rev.} \textbf{ D 49} (1994), no.~11, 6173.
  \href{http://dx.doi.org/10.1103/PhysRevD.49.6173}{\texttt{
  doi:10.1103/PhysRevD.49.6173}}.

\bibitem{ATLAS1}
\hrefCMSnoop {} {{ ATLAS} Collaboration, ``Search for Supersymmetry Using Final
  States with One Lepton, Jets, and Missing Transverse Momentum with the ATLAS
  Detector in $\sqrt{s}$ = 7 TeV $pp$ Collisions'',} \textit{ Phys. Rev. Lett.}
  \textbf{ 106} (2011) 131802.
  \href{http://dx.doi.org/10.1103/PhysRevLett.106.131802}{\texttt{
  doi:10.1103/PhysRevLett.106.131802}}.

\bibitem{SUS-10-003}
\hrefCMSnoop {} {{ CMS} Collaboration, ``Search for supersymmetry in pp
  collisions at 7 TeV in events with jets and missing transverse energy'',}
  \textit{ Phys. Lett.} \textbf{ B 698} (2011) 196.
  \href{http://dx.doi.org/10.1016/j.physletb.2011.03.021}{\texttt{
  doi:10.1016/j.physletb.2011.03.021}}.

\bibitem{SUS-10-007}
\hrefCMSnoop {} {{ CMS} Collaboration, ``Search for Physics Beyond the Standard
  Model in Opposite-sign Dilepton Events in pp Collisions at $\sqrt{s}$ = 7
  TeV'',}. \href{http://www.arXiv.org/abs/1103.1348}{\texttt{
  arXiv:1103.1348}}.

\bibitem{pythia}
\hrefCMSnoop {} {T.~Sj{\"o}strand, S.~Mrenna, and P.~Skands, ``PYTHIA 6.4
  Physics and Manual'',} \textit{ JHEP} \textbf{ 05} (2006) 026.

\bibitem{madgraph}
\hrefCMSnoop {} {J.~Alwall {et~al.}, ``MadGraph/MadEvent v4: The New Web
  Generation'',} \textit{ JHEP} \textbf{ 09} (2007) 028.
  \href{http://dx.doi.org/10.1088/1126-6708/2007/09/028}{\texttt{
  doi:10.1088/1126-6708/2007/09/028}}.

\bibitem{GEANT4}
\hrefCMSnoop {} {{ GEANT 4} Collaboration, ``GEANT4 -- a simulation toolkit'',}
  \textit{ Nucl. Instr. and Methods A} \textbf{ 506} (2003) 250.
  \href{http://dx.doi.org/10.1016/S0168-9002(03)01368-8}{\texttt{
  doi:10.1016/S0168-9002(03)01368-8}}.

\bibitem{ref:top}
\hrefCMSnoop {} {{ CMS} Collaboration, ``First Measurement of the Cross Section
  of Top-Quark Pair Production in Proton-Proton Collisions at $7$~TeV'',}
  \textit{ Phys. Lett.} \textbf{ B 695} (2011) 424.
  \href{http://dx.doi.org/10.1016/j.physletb.2010.11.058}{\texttt{
  doi:10.1016/j.physletb.2010.11.058}}.

\bibitem{ref:FR}
\href {http://cdsweb.cern.ch/record/1279147} {{ CMS} Collaboration,
  ``Performance of Methods for Data-Driven Background Estimation in {SUSY}
  Searches'',} \textit{ CMS Physics Analysis Summary} \textbf{
  CMS-PAS-SUS-10-001} (2010).

\bibitem{TRK-10-001}
\href {http://cdsweb.cern.ch/record/1258204} {{ CMS} Collaboration, ``Tracking
  and Vertexing Results from First Collisions'',} \textit{ CMS Physics Analysis
  Summary} \textbf{ CMS-PAS-TRK-10-001} (2010).

\bibitem{Adam:2005cg}
\href {http://cdsweb.cern.ch/record/934067} {W.~Adam {et~al.}, ``Track
  reconstruction in the CMS tracker'',} \textit{ CMS Note} \textbf{
  CMS-NOTE-2006-041}.

\bibitem{ref:GSF}
\hrefCMSnoop {} {W.~Adam {et~al.}, ``Reconstruction of electrons with the
  Gaussian-sum filter in the CMS tracker at the LHC'',} \textit{ J. Phys.}
  \textbf{ G 31} (2005) N9.
  \href{http://dx.doi.org/10.1088/0954-3899/31/9/N01}{\texttt{
  doi:10.1088/0954-3899/31/9/N01}}.

\bibitem{cteq66}
\hrefCMSnoop {} {H.-L. Lai {et~al.}, ``{Uncertainty induced by QCD coupling in
  the CTEQ-TEA global analysis of parton distributions}'',}
  \href{http://www.arXiv.org/abs/1004.4624}{\texttt{ arXiv:1004.4624}}.
  \href{http://dx.doi.org/10.1103/PhysRevD.82.054021}{\texttt{
  doi:10.1103/PhysRevD.82.054021}}.

\bibitem{lumi}
\href {http://cdsweb.cern.ch/record/1279145} {{ CMS} Collaboration,
  ``Measurement of {CMS} Luminosity'',} \textit{ CMS Physics Analysis Summary}
  \textbf{ CMS-PAS-EWK-10-004} (2010).

\bibitem{pdg}
\hrefCMSnoop {} {{ Particle Data Group} Collaboration, ``{Review of particle
  physics}'',} \textit{ J. Phys.} \textbf{ G 37} (2010) 075021.
\href{http://dx.doi.org/10.1088/0954-3899/37/7A/075021}{\texttt{
  doi:10.1088/0954-3899/37/7A/075021}}.

\bibitem{QCD-10-004pub}
\hrefCMSnoop {} {{ CMS} Collaboration, ``Charged particle multiplicities at
  $\sqrt{s}$=0.9, 2.36 and 7 TeV'',} \textit{ JHEP} \textbf{ 01} (2011) 079.
  \href{http://dx.doi.org/10.1007/JHEP01(2011)079}{\texttt{
  doi:10.1007/JHEP01(2011)079}}.

\bibitem{softsusy}
\hrefCMSnoop {} {B.~Allanach, ``SOFTSUSY: a program for calculating
  supersymmetric spectra'',} \textit{ Comput.Phys.Commun.} \textbf{ 143} (2002)
  305. \href{http://dx.doi.org/10.1016/S0010-4655(01)00460-X}{\texttt{
  doi:10.1016/S0010-4655(01)00460-X}}.

\bibitem{sdecay}
\hrefCMSnoop {} {M.~Muhlleitner, A.~Djouadi, and Y.~Mambrini, ``SDECAY: a
  Fortran code for the decays of the supersymmetric particles in the MSSM'',}
  \textit{ Comput. Phys. Commun.} \textbf{ 168} (2005) 46.
  \href{http://dx.doi.org/10.1016/j.cpc.2005.01.012}{\texttt{
  doi:10.1016/j.cpc.2005.01.012}}.

\bibitem{cteq}
\hrefCMSnoop {} {P.~Nadolsky {et~al.}, ``Implications of CTEQ global analysis
  for collider observables'',} \textit{ Phys. Rev.} \textbf{ D 78} (2008)
  013004. \href{http://dx.doi.org/10.1103/PhysRevD.78.013004}{\texttt{
  doi:10.1103/PhysRevD.78.013004}}.

\bibitem{prospino}
\hrefCMSnoop {} {W.~Beenakker {et~al.}, ``Squark and Gluino Production at
  Hadron Colliders'',} \textit{ Nucl. Phys.} \textbf{ B 492} (1997) 51.
  \href{http://dx.doi.org/10.1016/S0550-3213(97)80027-2}{\texttt{
  doi:10.1016/S0550-3213(97)80027-2}}.

\bibitem{CDFtrilep}
\hrefCMSnoop {} {{ CDF} Collaboration, ``Search for Supersymmetry in $p\bar{p}$
  Collisions at $\sqrt{s} = 1.96$ TeV using the Trilepton Signature for
  Chargino-Neutralino Production'',} \textit{ Phys. Rev. Lett.} \textbf{ 101}
  (2008) 251801.
  \href{http://dx.doi.org/10.1103/PhysRevLett.101.251801}{\texttt{
  doi:10.1103/PhysRevLett.101.251801}}.

\bibitem{Abazov200934}
\hrefCMSnoop {} {{ D0} Collaboration, ``Search for associated production of
  charginos and neutralinos in the trilepton final state using 2.3~fb$^{-1}$ of
  data'',} \textit{ Phys. Lett.} \textbf{ B 680} (2009) 34.
  \href{http://dx.doi.org/10.1016/j.physletb.2009.08.011}{\texttt{
  doi:10.1016/j.physletb.2009.08.011}}.

\bibitem{leplimit}
\hrefCMSnoop {} {{ ALEPH, DELPHI, L3 and OPAL} Collaboration, ``Joint SUSY
  Working Group'',} \textit{ LEPSUSYWG} \textbf{ 02-06-2}.

\bibitem{aleph}
\hrefCMSnoop {} {{ ALEPH} Collaboration, ``Absolute mass lower limit for the
  lightest neutralino of the MSSM from $e^+e^-$ data at $\sqrt{s}$ up to 209
  GeV'',} \textit{ Phys. Lett.} \textbf{ B 583} (2004) 247. See also references
  therein.

\bibitem{delphi}
\hrefCMSnoop {} {{ DELPHI} Collaboration, ``Searches for supersymmetric
  particles in $e^+e^-$ collisions up to 208 GeV and interpretation of the
  results within the MSSM'',} \textit{ Eur. Phys. J.} \textbf{ C 31} (2003)
  421. See also references therein.

\bibitem{l3}
\hrefCMSnoop {} {{ L3} Collaboration, ``Search for scalar leptons and scalar
  quarks at LEP'',} \textit{ Phys. Lett.} \textbf{ B 580} (2004) 37. See also
  references therein.

\bibitem{opal}
\hrefCMSnoop {} {{ OPAL} Collaboration, ``Search for chargino and neutralino
  production at $\sqrt{s} = 192-209$ GeV at LEP'',} \textit{ Eur. Phys. J.}
  \textbf{ C 35} (2004) 1. See also references therein.

\end{thebibliography}\endgroup

\cleardoublepage\appendix\section{The CMS Collaboration \label{app:collab}}\begin{sloppypar}\hyphenpenalty=5000\widowpenalty=500\clubpenalty=5000\textbf{Yerevan Physics Institute,  Yerevan,  Armenia}\\*[0pt]
S.~Chatrchyan, V.~Khachatryan, A.M.~Sirunyan, A.~Tumasyan
\vskip\cmsinstskip
\textbf{Institut f\"{u}r Hochenergiephysik der OeAW,  Wien,  Austria}\\*[0pt]
W.~Adam, T.~Bergauer, M.~Dragicevic, J.~Er\"{o}, C.~Fabjan, M.~Friedl, R.~Fr\"{u}hwirth, V.M.~Ghete, J.~Hammer\cmsAuthorMark{1}, S.~H\"{a}nsel, M.~Hoch, N.~H\"{o}rmann, J.~Hrubec, M.~Jeitler, G.~Kasieczka, W.~Kiesenhofer, M.~Krammer, D.~Liko, I.~Mikulec, M.~Pernicka, H.~Rohringer, R.~Sch\"{o}fbeck, J.~Strauss, F.~Teischinger, P.~Wagner, W.~Waltenberger, G.~Walzel, E.~Widl, C.-E.~Wulz
% ...
\vskip\cmsinstskip
% ...
1:~~Also at CERN, European Organization for Nuclear Research, Geneva, Switzerland\\
% ...

\end{sloppypar}
\end{document}